\documentclass[structabstract]{aaold}
\usepackage[utf8]{inputenc}
\usepackage{txfonts}
\usepackage{graphicx}
\usepackage{natbib}
\usepackage{multirow}
\usepackage{soul}
\usepackage{upgreek}

\usepackage{amsmath}
\usepackage{url}
\usepackage{hyperref}
\usepackage{placeins}
\usepackage{afterpage}
\hypersetup{
        bookmarksnumbered = true,
        colorlinks = true,
        unicode = true,
        urlcolor = blue,
        linkcolor = blue,
        citecolor = blue
        }

\bibpunct{(}{)}{;}{a}{}{,}

\begin{document}

\newcommand{\ie}{{\em i.e.,}}
\newcommand{\eg}{{\em e.g.,}}

\newcommand{\emm}[1]{\ensuremath{#1}}
\newcommand{\emr}[1]{\emm{\mathrm{#1}}}
\newcommand{\chem}[1]{\emr{\,#1}}
\newcommand{\unit}[1]{\emr{\,#1}}
\newcommand{\e}[1]{\emm{\times 10^{#1}}}
\newcommand{\diff}{\emr{d}}

\newcommand{\Av}{\emm{A_\emr{V}}}
\newcommand{\Ak}{\emm{A_\emr{K}}}
\newcommand{\Msun}{\emm{M_{\odot}}}

\newcommand{\GB}{\unit{GB}}
\newcommand{\kHz}{\unit{kHz}}
\newcommand{\MHz}{\unit{MHz}}
\newcommand{\GHz}{\unit{GHz}}
\newcommand{\Myr}{\unit{Myr}}
\newcommand{\pc}{\unit{pc}}
\newcommand{\mpc}{\unit{mpc}}
\newcommand{\pscm}{\unit{cm^{-2}}} 
\newcommand{\pccm}{\unit{cm^{-3}}} 
\newcommand{\Htpscm}{\unit{H_2\,cm^{-2}}} 
\newcommand{\Htpccm}{\unit{H_2\,cm^{-2}}} 
\newcommand{\magn}{\unit{mag}}
\newcommand{\micron}{\unit{\upmu m}}
\newcommand{\K}{\unit{K}}
\newcommand{\mm}{\unit{mm}}
\newcommand{\kms}{\unit{km\,s^{-1}}}

\newcommand{\Hii}{\textsc{Hii}}
\newcommand{\SNR}{signal-to-noise ratio}
\newcommand{\Nh}{\emm{N_\emr{H}}}
\newcommand{\Nhtwo}{\emm{N_\emr{H_2}}}
\newcommand{\rhoh}{\emm{\rho_\emr{H}}}
\newcommand{\rhohhtwo}{\emm{\rho_\emr{H_2}}}

\newcommand{\CeO}{\emr{C}\emm{^{18}}\emr{O}}
\newcommand{\CeOline}{\CeO{}\emm{\,(J=1-0)}}
\newcommand{\tCO}{\emm{^{13}}\emr{CO}}
\newcommand{\tCOline}{\tCO{}\emm{\,(J=1-0)}}
\newcommand{\CO}{\emm{^{12}}\emr{CO}}
\newcommand{\COline}{\CO{}\emm{\,(J=1-0)}}
\newcommand{\HCOp}{\emr{HCO}\emm{^{+}}}
\newcommand{\HCOpline}{\HCOp{}\emm{\,(J=1-0)}}

\newcommand{\GMC}{Giant Molecular Cloud}
\newcommand{\coltt}{\texttt{Colume}}


\newcommand{\FigToyGeometry}{
    \begin{figure}
        \centering
        \includegraphics[width=\linewidth]{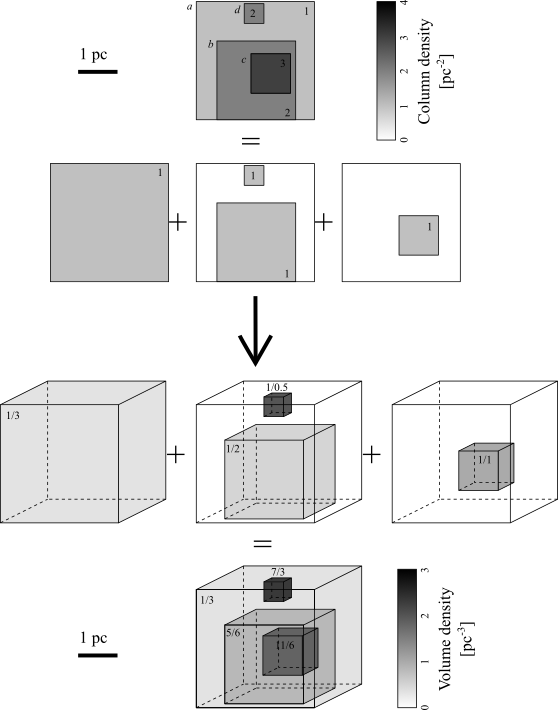}
        \caption{Schematic representation of the volume density estimation. The indicated numbers correspond to the column (respectively volume) density values, in arbitrary units. \emph{Top half:} Original column density map with contours that define the following structures: a 9\pc$^2$ diffuse background \emph{(a)}, a main cloud with a 4\pc$^2$ intermediate density envelope \emph{(b)} and a 1\pc$^2$ dense core \emph{(c)}, and a small isolated 0.25\pc$^2$ dense clump \emph{(d)}. The column density map is then decomposed into column density increments. \emph{Bottom half:} The column density increments are converted into volume density increments by ascribing to them depths equal to the square root of their area (3\pc, 2\pc, 1\pc{} and 0.5\pc{} for the diffuse background, the envelope, the core and the clump respectively). The final volume density structure is obtained by summing the volume density increments, assuming that each successive increment is nested within the volume underlying the previous contour. Note that the nested volumes are represented in the mid-plane along the line of sight only by convention.}
        \label{fig:toygeometry}
    \end{figure}
}

\newcommand{\FigToyPDF}{
    \begin{figure}
        \centering
        \includegraphics[width=\linewidth]{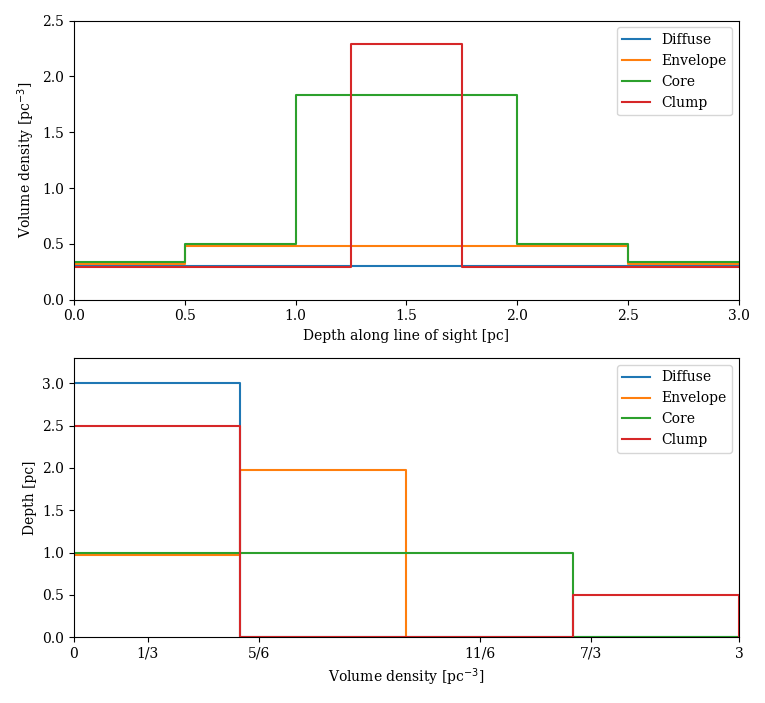}
        \caption{\emph{Top:} Density profile along lines of sight corresponding to the different environments in the map (diffuse background, intermediate density envelope, dense core, small isolated clump) described in Fig. \ref{fig:toygeometry}. Note that the nested dense structures are represented in the mid-plane along the line of sight only by convention. \emph{Bottom:} PDFs of volume densities along the line of sight corresponding to the above lines of sight. The densities are expressed in the same arbitrary units as in Fig. \ref{fig:toygeometry}. The statistical weights are equivalent to depths along the line of sight, as represented in the top panel.}
        \label{fig:toypdf}
    \end{figure}
}

\newcommand{\FigToyPanels}{
    \begin{figure}
        \centering
        \includegraphics[width=\linewidth]{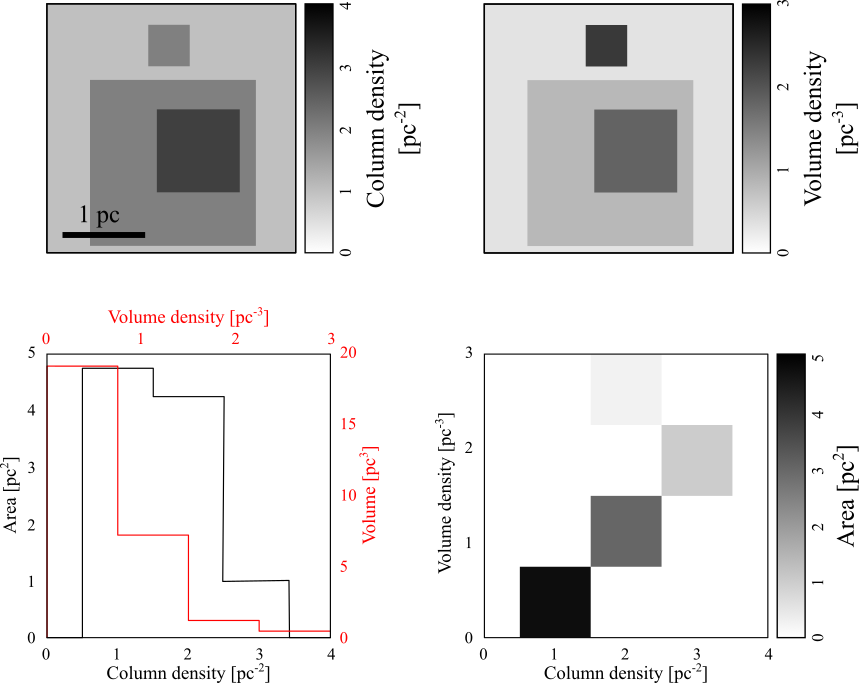}
        \caption{The main products of the \coltt{} volume density estimation (similar to Fig. \ref{fig:panels}) obtained for the simple example illustrated in Fig. \ref{fig:toygeometry}. \emph{Top left:} Original column density data. \emph{Top right:} Peak volume density reached along each line of sight. \emph{Bottom left:} Comparison of the column density (black) and volume density (red) histograms for the entire area (respectively volume) of the cloud. \emph{Bottom right:} Joint distribution of the column density and peak volume density.}
        \label{fig:toypanels}
    \end{figure}
}

\newcommand{\FigPDFs}{
    \begin{figure}
        \centering
        \includegraphics[width=\linewidth]{figures/pdfs/pdfs2-Cha II.png}
        \caption{Comparison of the column density PDF (solid red) and reconstructed volume density PDF (solid red) in the case of the Cha II cloud. The other clouds are shown in Appendix \ref{app:voldens}. The dash-dotted line corresponds to the smoothed volume density PDF used to determine the volume density contrast. Note that the data range are different but the scaling is the same for the column and volume density case.}
        \label{fig:pdfs}
    \end{figure}
}

\newcommand{\FigPanels}{
    \begin{figure}
        \centering
        \includegraphics[width=\linewidth]{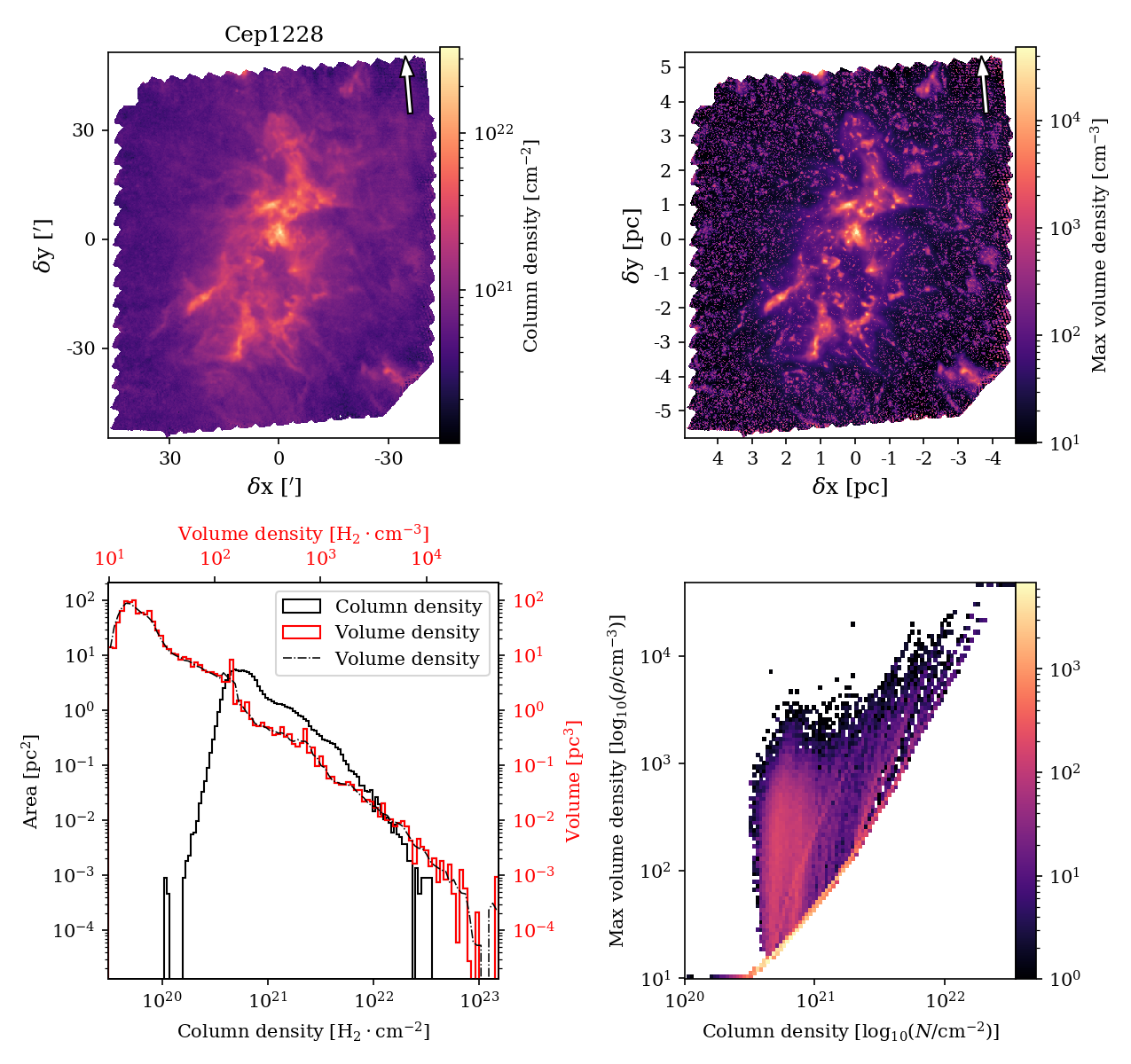}
        \caption{The main products of the \coltt{} volume density estimation for the Cep1228 molecular cloud. The same figures for all clouds in our sample are shown in the \href{https://zenodo.org/records/14360623}{supplementary online material}. \emph{Top left:} Column density map obtained from HGBS data, the arrow indicates North. \emph{Top right:} Map of the peak volume density (maximal volume density reached along a line of sight). \emph{Bottom left:} 
        Comparison of the column density PDF (solid black) and reconstructed volume density PDF (solid red) for the entire cloud. The dash-dotted line corresponds to the smoothed volume density PDF used to determine the volume density contrast. Note that the data range are different but the scaling is the same for the column and volume density case.
        \emph{Bottom right:} Joint distribution of the column density and peak volume density for each line of sight.}
        \label{fig:panels}
    \end{figure}
}

\newcommand{\FigColVol}{
    \begin{figure}
        \centering
        \includegraphics[width=\linewidth]{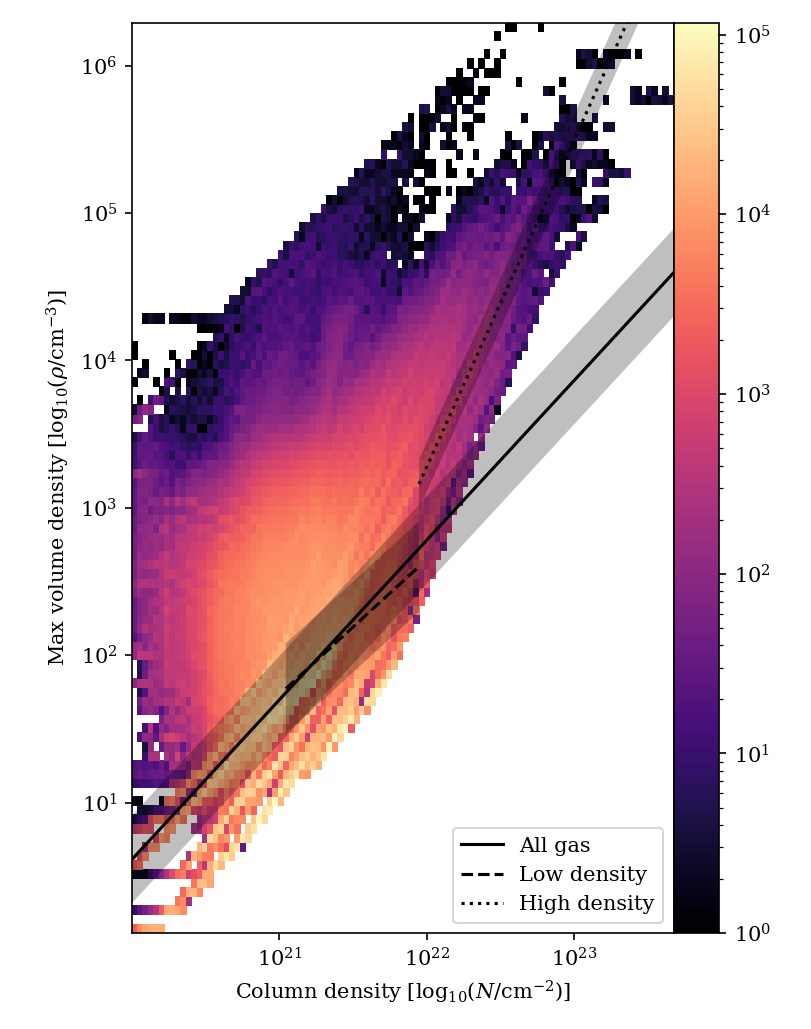}
        \caption{Joint distribution of the column density and peak volume density, for all the lines of sight of all the clouds in the sample. The trend of this distribution has been fitted as a power-law in three column density regimes: for the entire column density range present in the data (solid line), for the ``molecular low density gas'' defined as 1 -- 8 \Av{} (dashed line), and for the ``dense gas'' defined as $\Av > 8$ (dotted line). The RMS scatter around the best fit relation is represented by the shaded areas.}
        \label{fig:colvol}
    \end{figure}
}

\newcommand{\FigSFE}{
    \begin{figure*}
        \centering
        \includegraphics[width=\linewidth]{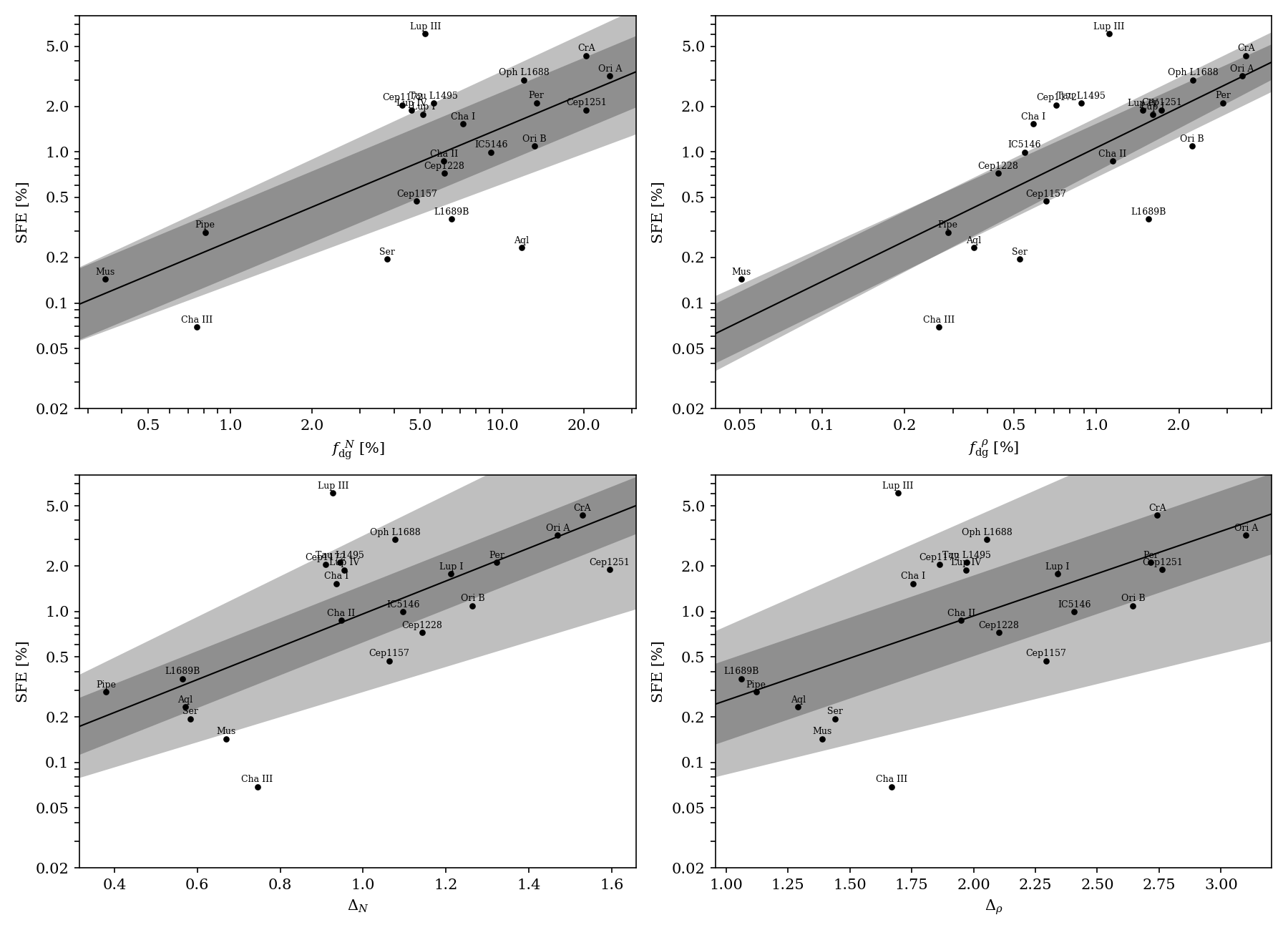}
        \caption{Correlation of SFE with various metrics of the dense gas component in clouds. In each case a power-law is fitted (solid line), the fit errors are represented in light grey, and the (Gaussian) spread of the cloud of points with respect to the fitted line is represented in dark grey. \emph{Top left:} SFE vs. column-density based dense gas fraction. \emph{Top right:} SFE vs. volume-density based dense gas fraction. \emph{Bottom left:} SFE vs. column-density based density contrast. \emph{Bottom right:} SFE vs. volume-density based density contrast.}
        \label{fig:sfe}
    \end{figure*}
}

\newcommand{\FigScatter}{
    \begin{figure}
        \centering
        \includegraphics[width=\linewidth]{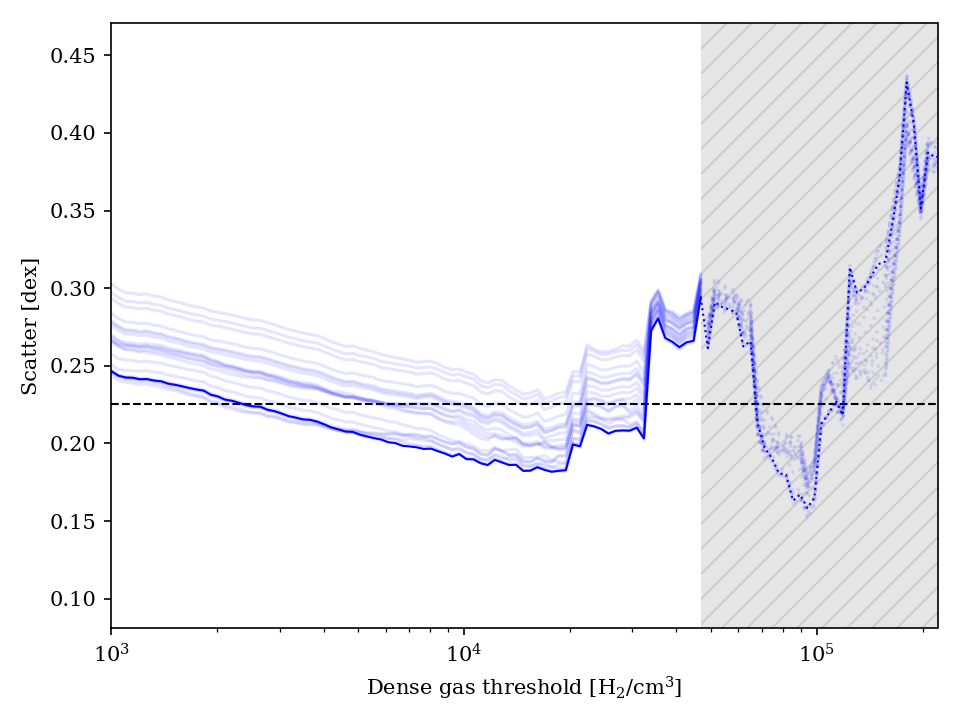}
        \caption{Determination of the volume density thresholds yielding the tightest correlation between the volume density and star formation. Each transparent blue line corresponds to a different value of the low density threshold between $10\Htpccm$ and $5\e2\unit{\emr{H}_2\pccm}$, the solid blue plot corresponds to the optimal threshold of $2\e2 \Htpccm$. The dashed horizontal line at 0.225 dex corresponds to the spread obtained when using column densities. Values for a dense gas threshold higher than 5\e4\,H$_2$\pccm{} (shaded area) are not considered, because not all clouds in the sample reach such high values and the statistics become thus less and less reliable.}
        \label{fig:scatter}
    \end{figure}
}

\newcommand{\FigWidth}{
    \begin{figure}
        \centering
        \includegraphics[width=\linewidth]{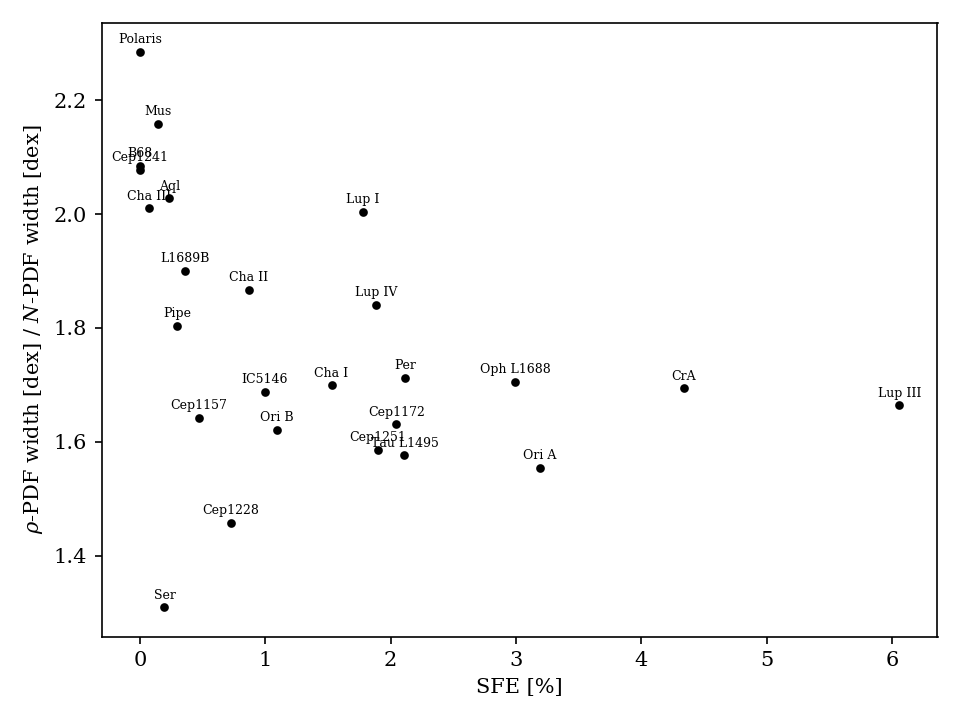}
        \caption{Relation between the star formation efficiency of molecular clouds and the ratio between the widths (in dex) of their volume density PDF and column density PDF.}
        \label{fig:width}
    \end{figure}
}

\newcommand{\FigResolution}{
    \begin{figure}
        \centering
        \includegraphics[width=\linewidth]{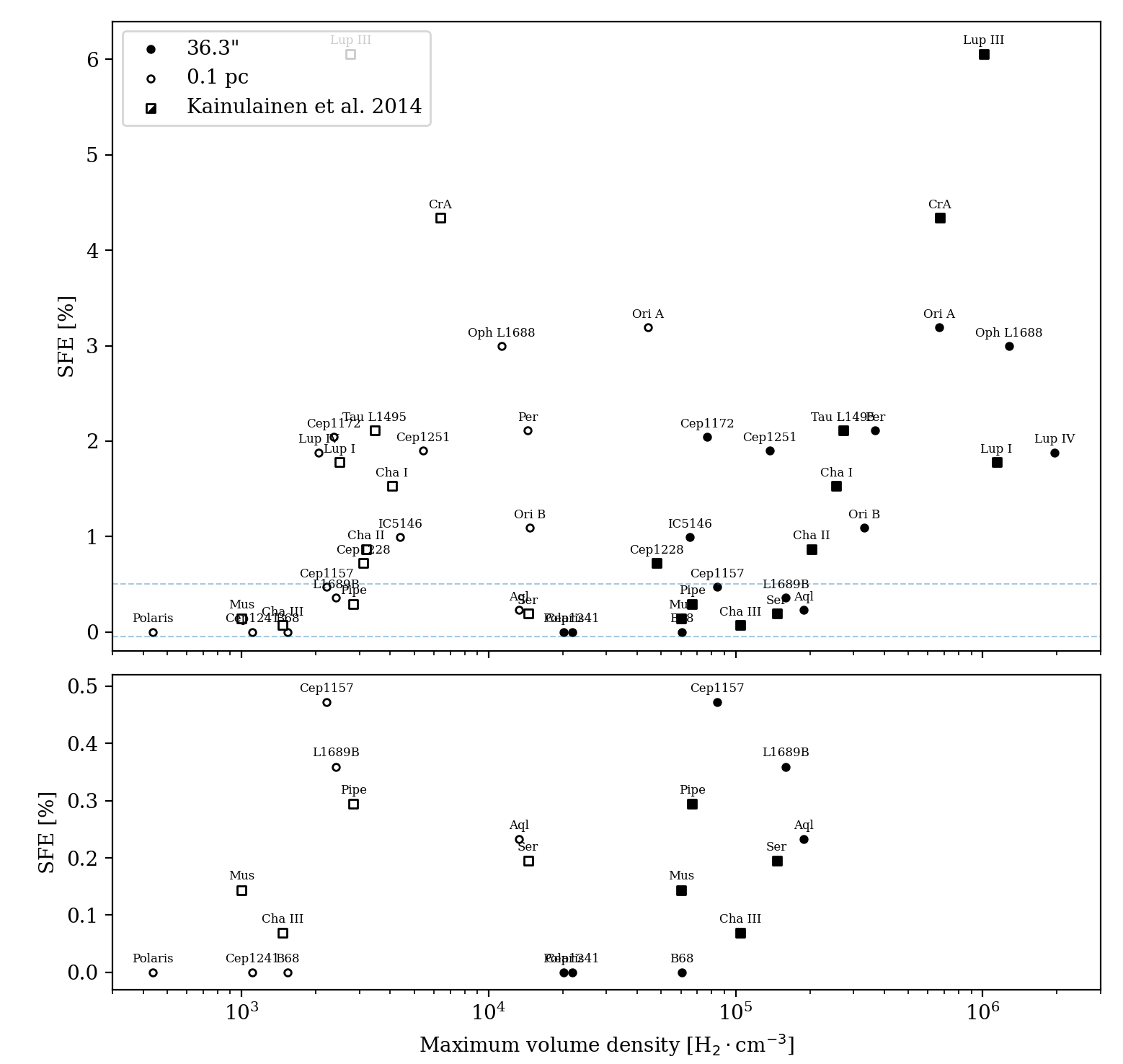}
        \caption{\emph{Top:} Comparison between the maximum volume densities obtained for the clouds in our sample at the fixed angular resolution of 36.3'' provided by the \emph{Herschel} telescope, and at the fixed spatial resolution of 0.1\pc{} used by \citet{kainulainen14}. The 12 clouds in common between the \citet{kainulainen14} sample and ours are highlighted by square markers. \emph{Bottom:} Zoom on the area defined by the horizontal, dashed lines in the top panel, at the very low SFE relevant for defining a star formation threshold.}
        \label{fig:resolution}
    \end{figure}
}

\newcommand{\FigAppPdf}{
    \begin{figure}
        \centering
        \includegraphics[width=\linewidth]{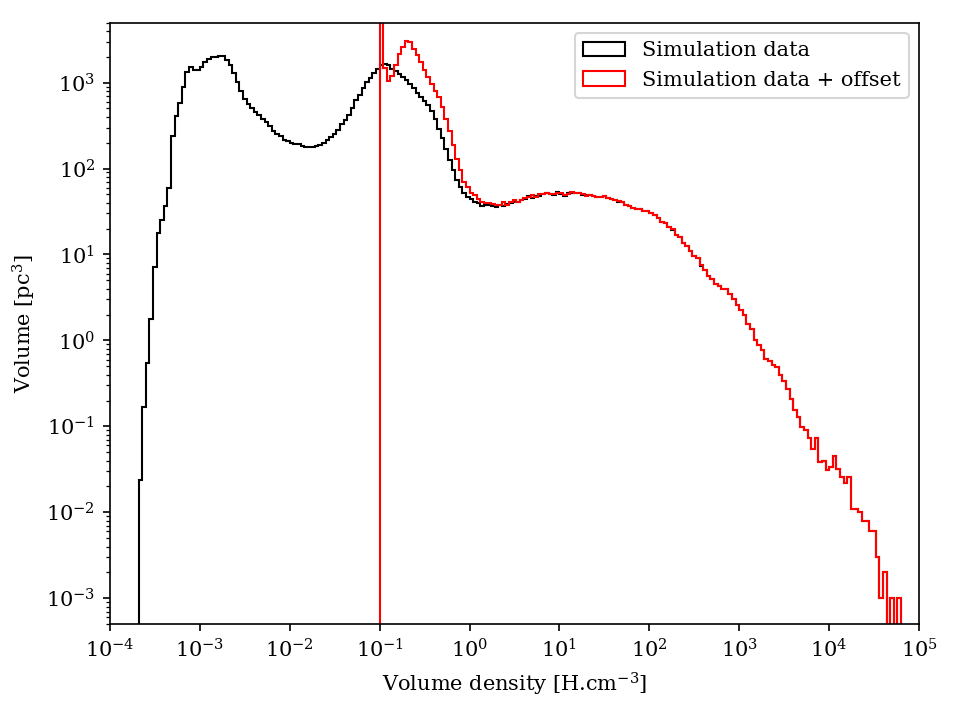}
        \caption{Volume density distribution in the simulated test cloud, which has an initial mass $M = 4\e3 \Msun$ and lies in a $40\times40\times40\pc$ cube. An offset density is added to the cube to simulate a diffuse fore- and background when simulating column density observations.}
        \label{fig:app:pdf}
    \end{figure}
}

\newcommand{\FigAppCleanM}{
    \begin{figure}
        \centering
        \includegraphics[width=\linewidth]{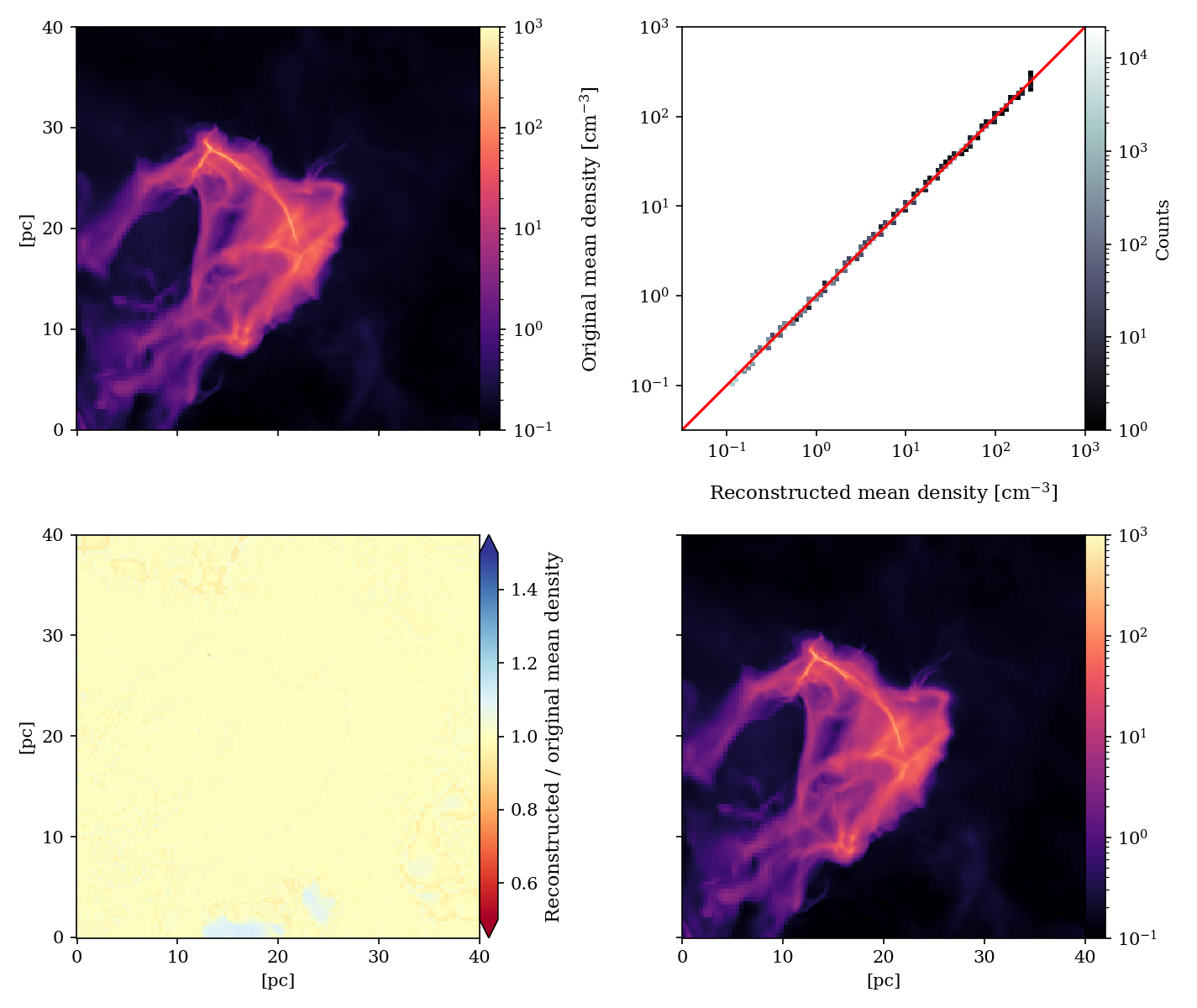}
        \caption{Estimation of the mean volume density along the line of sight from a noiseless column density map. \emph{Top left:} Original noiseless data. \emph{Top right:} Joint histogram of the original vs. reconstructed mean volume densities. \emph{Bottom left:} Map of the ratio of reconstructed-to-original mean volume densities. \emph{Bottom right:} Reconstructed map of mean volume densities along the line of sight.}
        \label{fig:app:cleanm}
    \end{figure}
}

\newcommand{\FigAppCleanP}{
    \begin{figure}
        \centering
        \includegraphics[width=\linewidth]{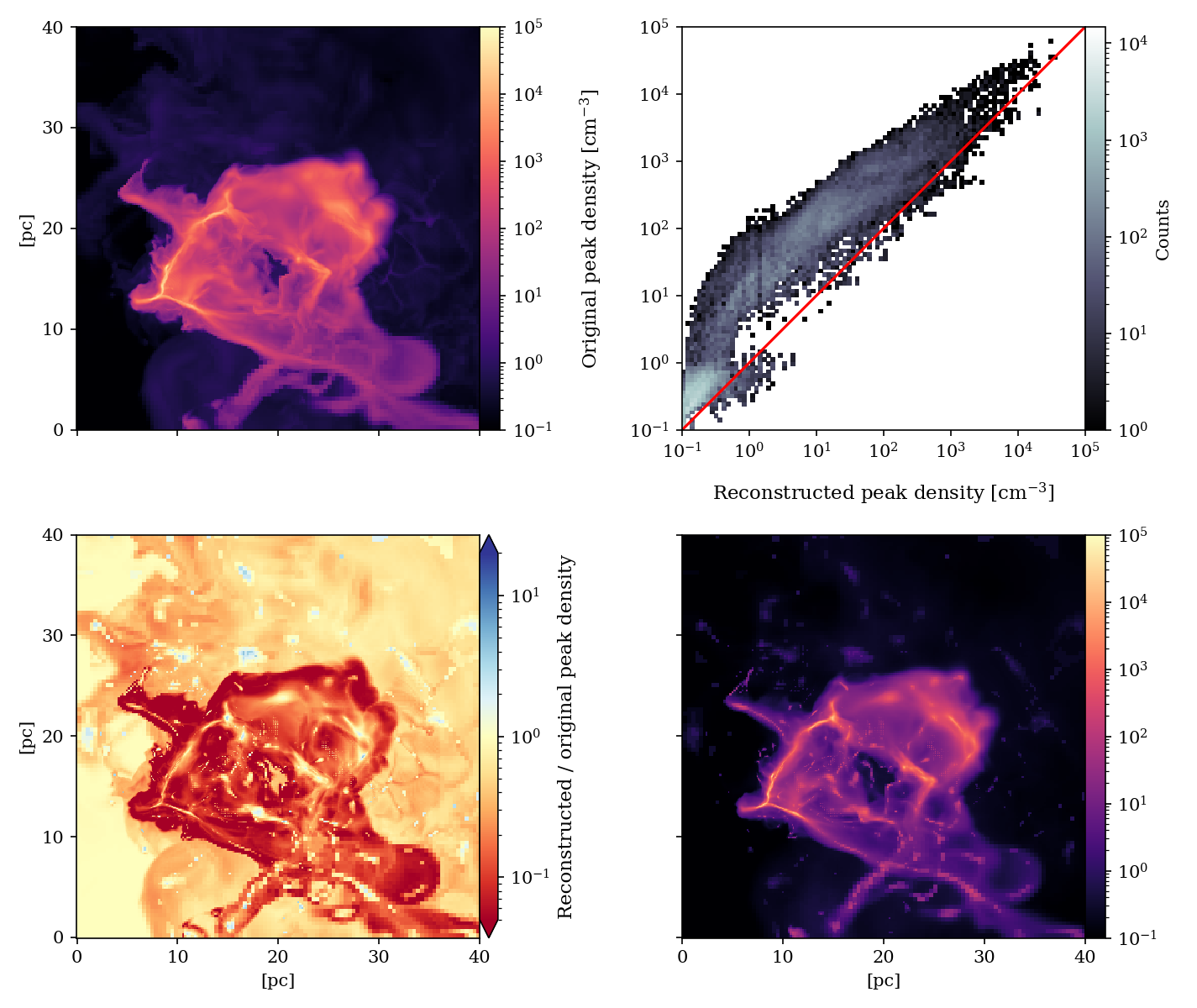}
        \caption{Estimation of the peak volume density along the line of sight from a noiseless column density map. \emph{Top left:} Original noiseless data. \emph{Top right:} Joint histogram of the original vs. reconstructed peak volume densities. \emph{Bottom left:} Map of the ratio of reconstructed-to-original peak volume densities. \emph{Bottom right:} Reconstructed map of peak volume densities along the line of sight.}
        \label{fig:app:cleanp}
    \end{figure}
}

\newcommand{\FigAppPDFxyz}{
    \begin{figure}
        \centering
        \includegraphics[width=\linewidth]{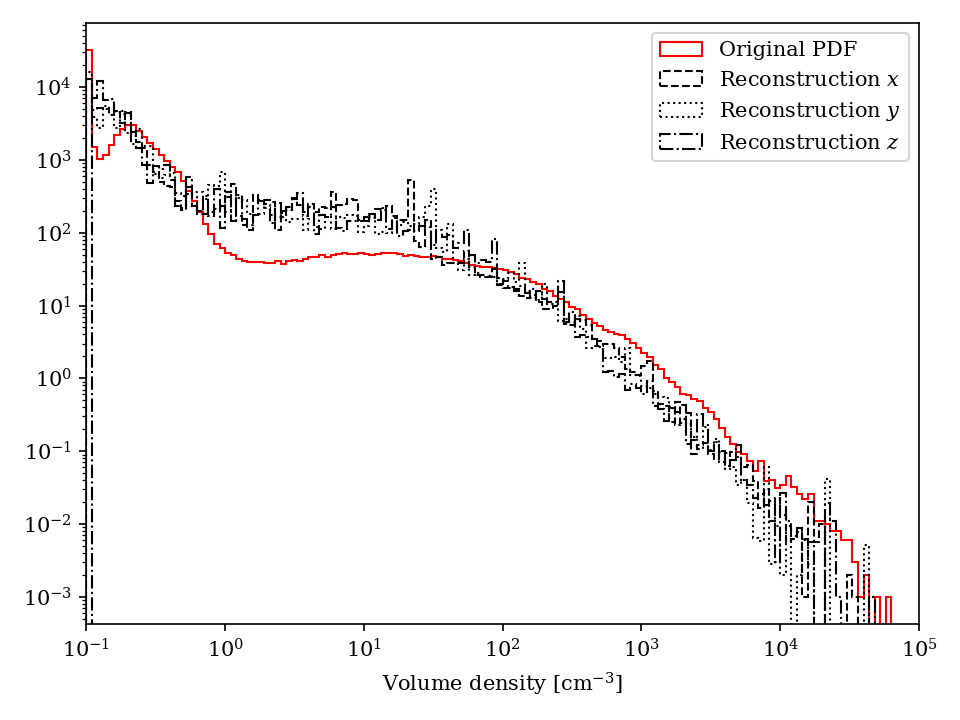}
        \caption{True volume density distribution of the simulated data cube (Fig. \ref{fig:app:pdf}), compared to the reconstructed distributions based on the analysis of the projected column density maps.}
        \label{fig:app:pdfxyz}
    \end{figure}
}

\newcommand{\FigAppNoiseOne}{
    \begin{figure}[tp]
        \centering
        \includegraphics[width=\linewidth]{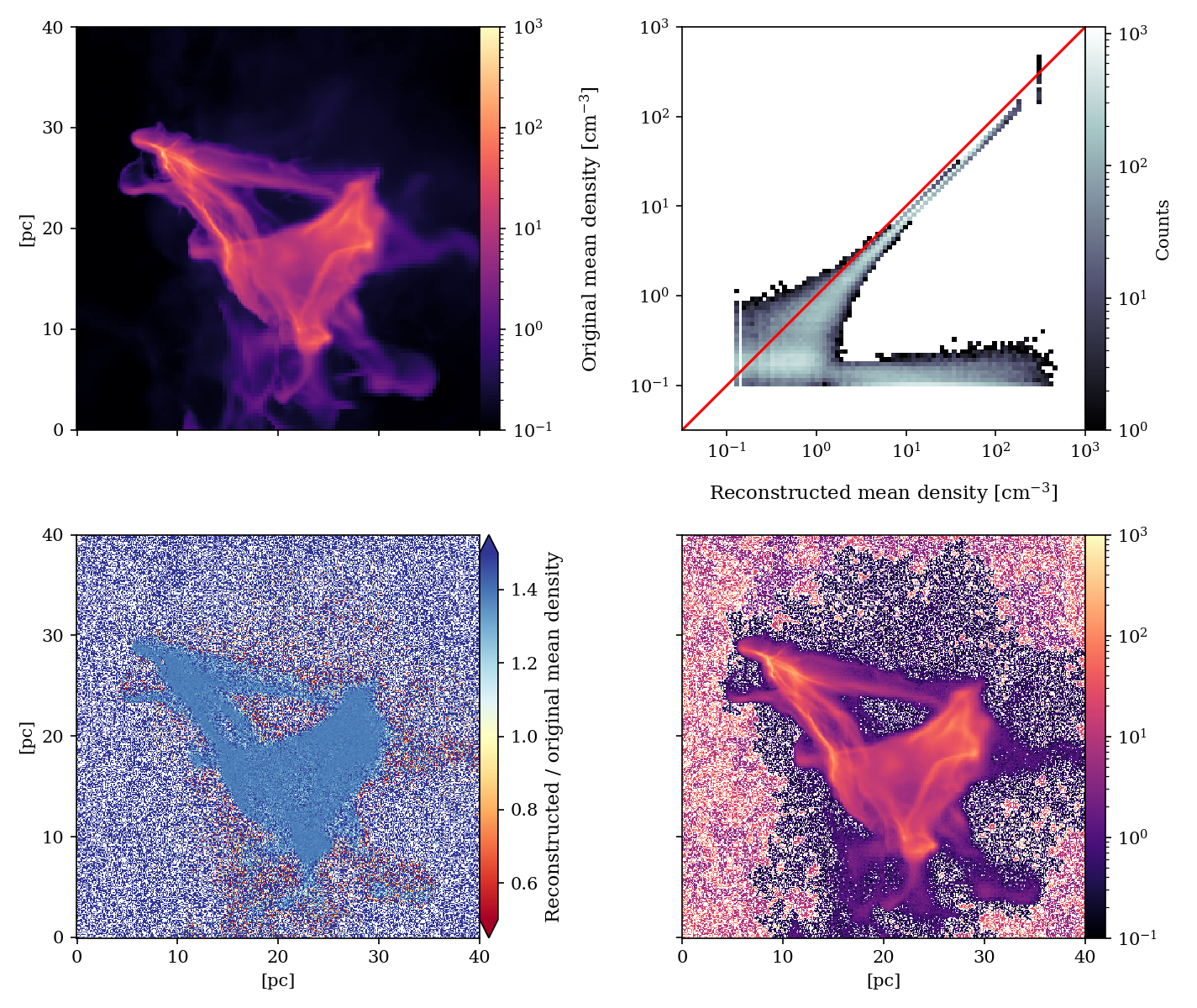}
        \includegraphics[width=\linewidth]{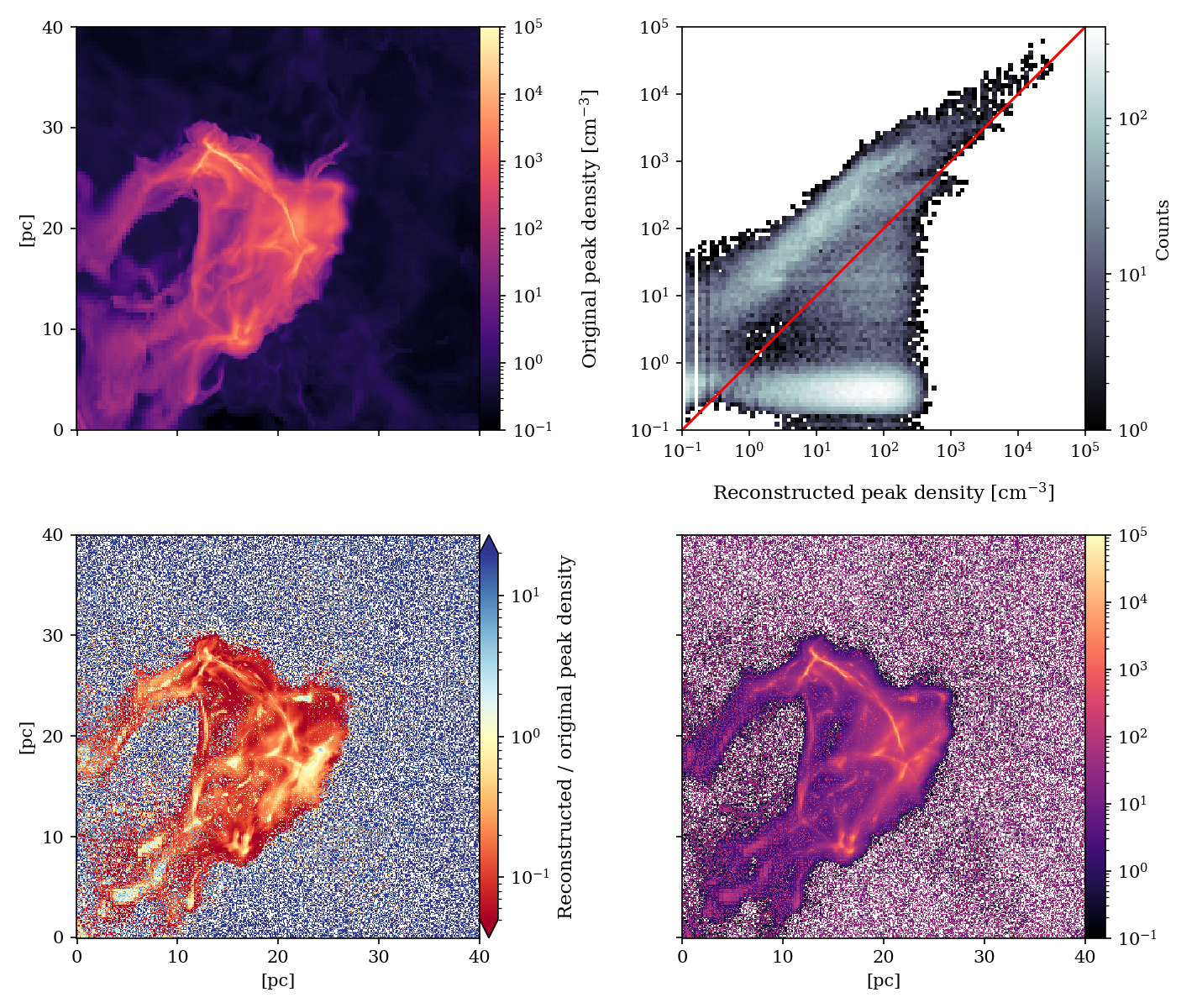}
        \includegraphics[width=\linewidth]{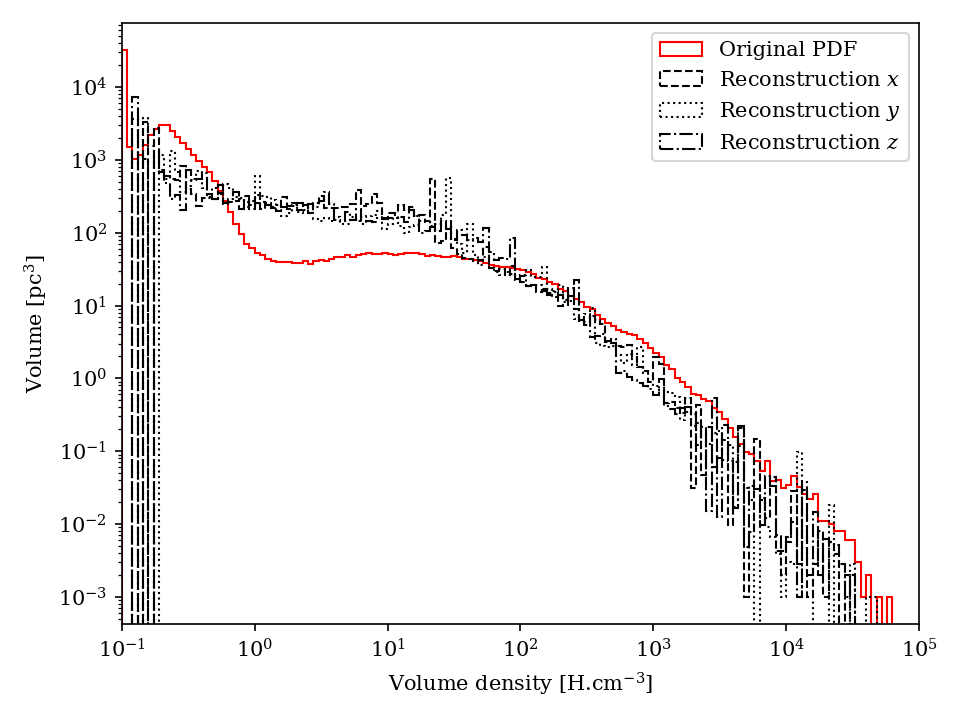}
        \caption{\emph{Top four panels:} Same as Fig. \ref{fig:app:cleanp}, but for the mean volume density along the line of sight, for a different projection, and with added uniform noise. \emph{Middle four panels:} Same as above, but for the peak volume density along the line of sight, and for a different projection. \emph{Bottom:} Same as Fig. \ref{fig:app:pdfxyz}, but with added uniform noise.}
        \label{fig:app:noiseone}
    \end{figure}
}

\newcommand{\FigAppNoiseTwo}{
    \begin{figure}[tp]
        \centering
        \includegraphics[width=\linewidth]{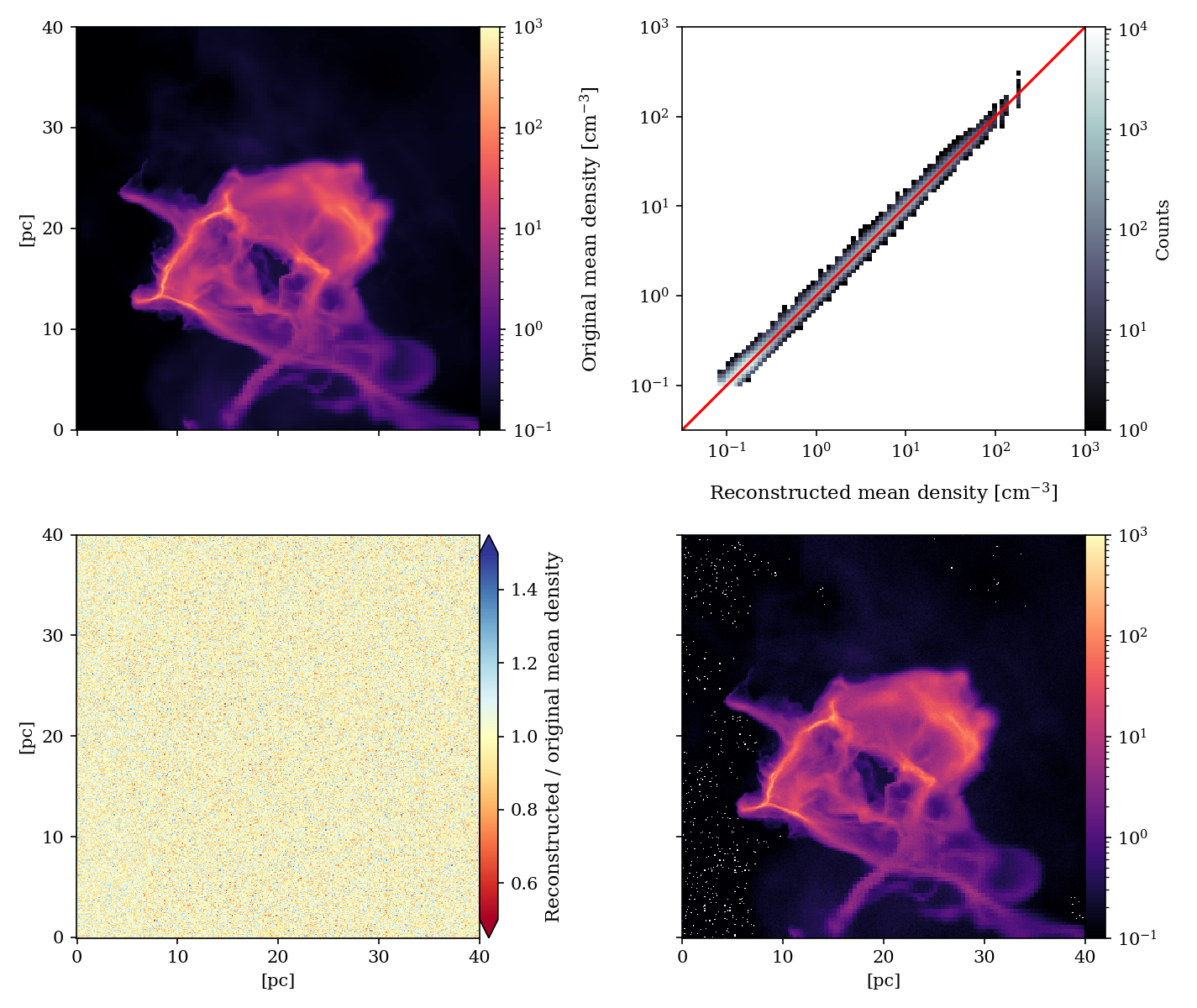}
        \includegraphics[width=\linewidth]{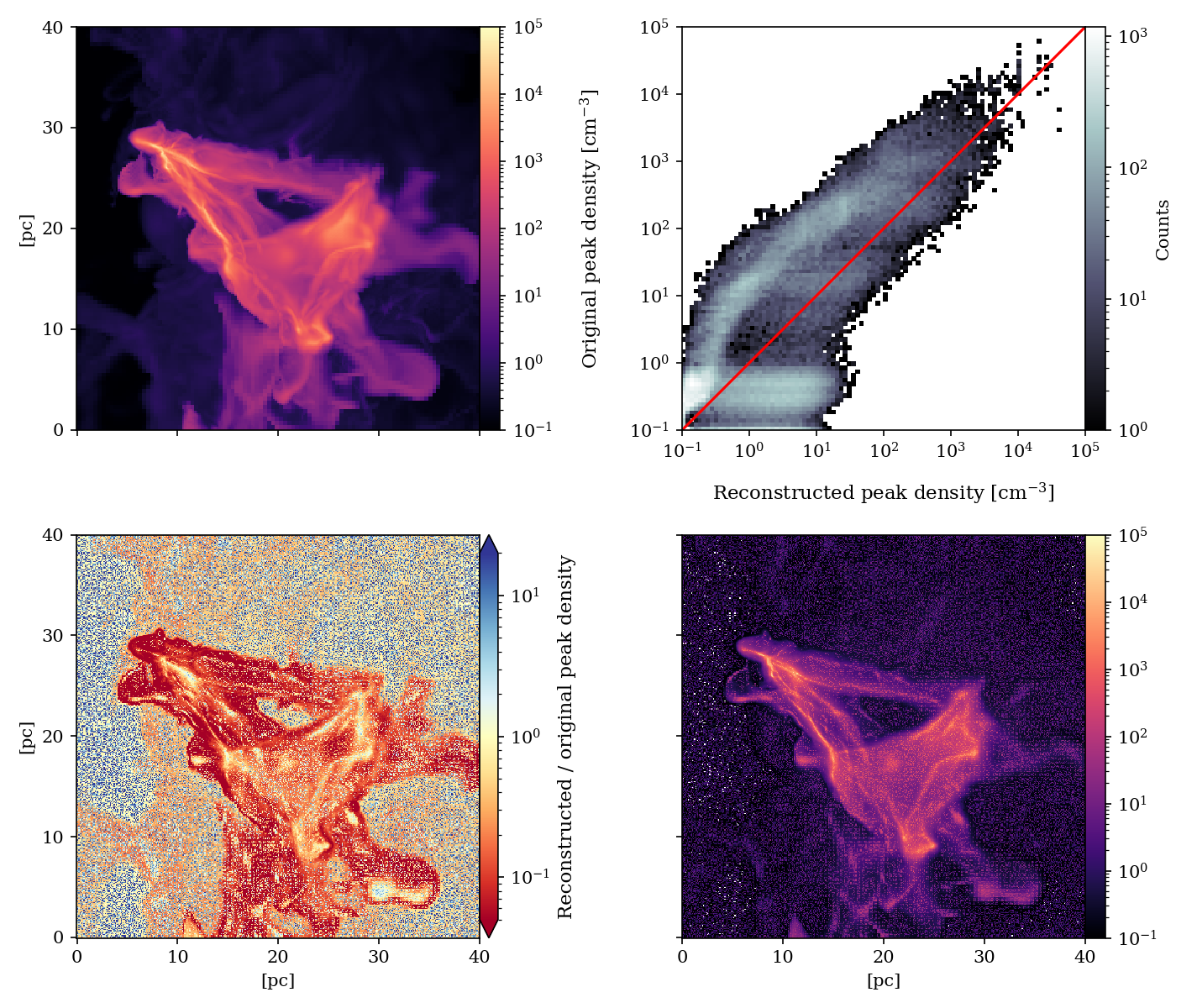}
        \includegraphics[width=\linewidth]{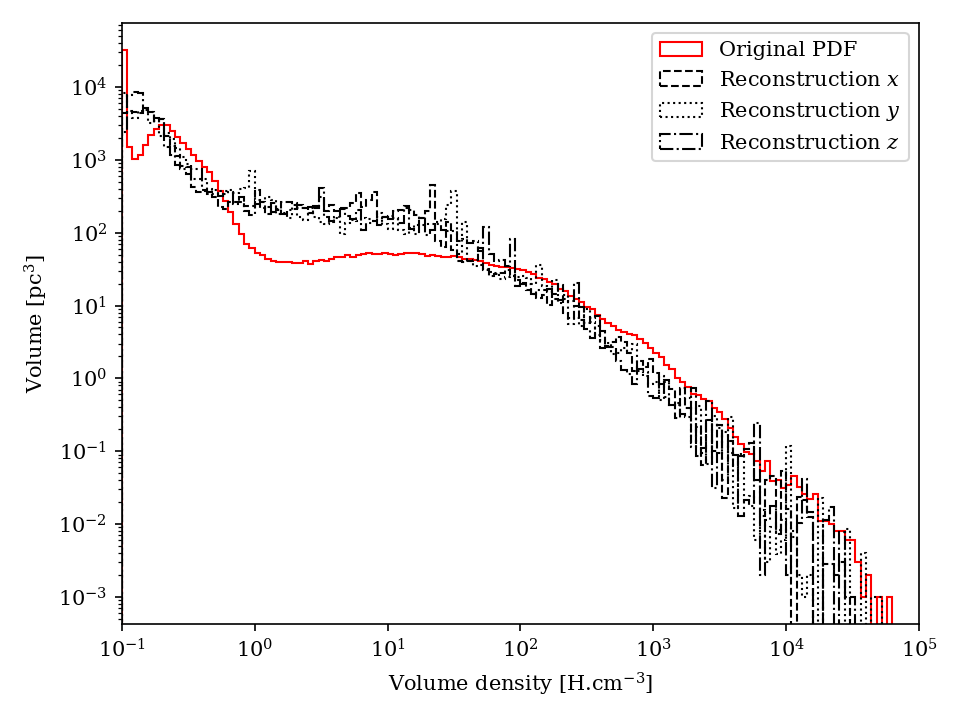}
        \caption{Same as Fig. \ref{fig:app:noiseone}, but with a uniform \SNR\ and different projections.}
        \label{fig:app:noisetwo}
    \end{figure}
}

\newcommand{\FigAppSamplingCol}{
    \begin{figure}
        \centering
        \includegraphics[width=0.49\linewidth]{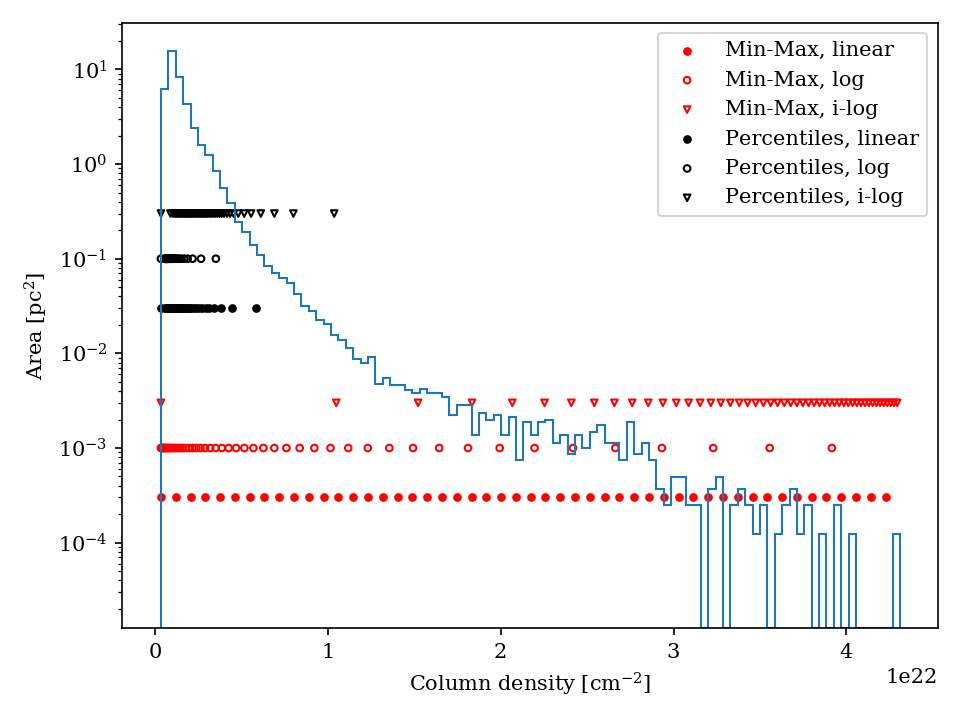}
        \includegraphics[width=0.49\linewidth]{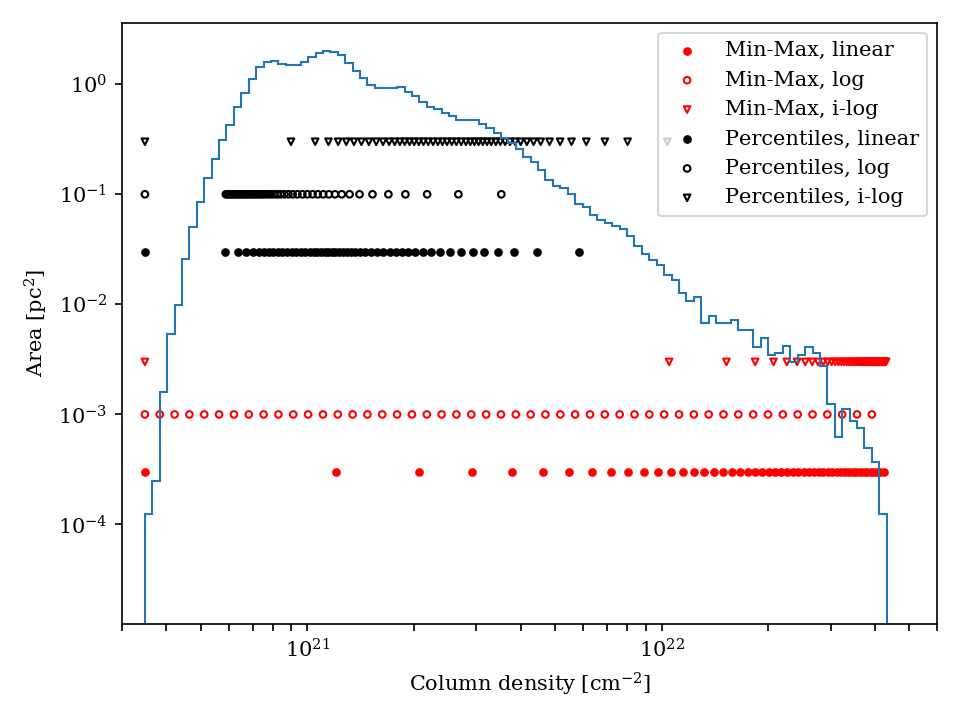}\\
        \caption{Comparison of the bin locations for the various sampling options implemented in \coltt{} applied to the column density distribution of the Cha II cloud, with a coarse 50-level sampling, in linear and logarithmic scales.}
        \label{fig:app:samplingcol}
    \end{figure}
}

\newcommand{\FigAppSamplingPDF}{
    \begin{figure}
        \centering
        \includegraphics[width=\linewidth]{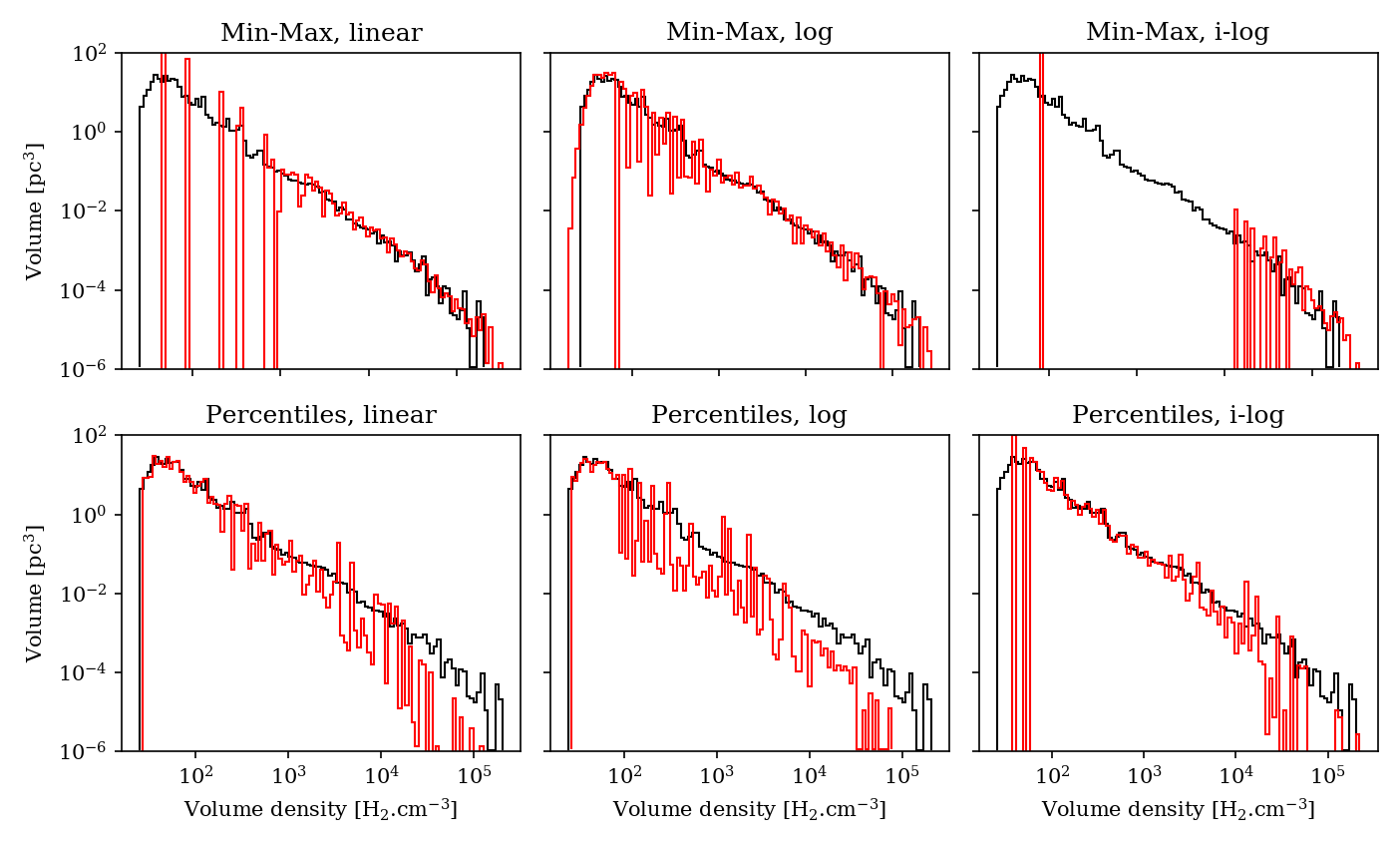}
        \caption{Effects of sampling options on the global volume density PDF, illustrated by comparing coarse, 50-level volume density reconstructions with a 2000-level benchmark, for the Cha II cloud.}
        \label{fig:app:samplingpdf}
    \end{figure}
}

\newcommand{\FigAppSamplingMaps}{
    \begin{figure}
        \centering
        \includegraphics[width=\linewidth]{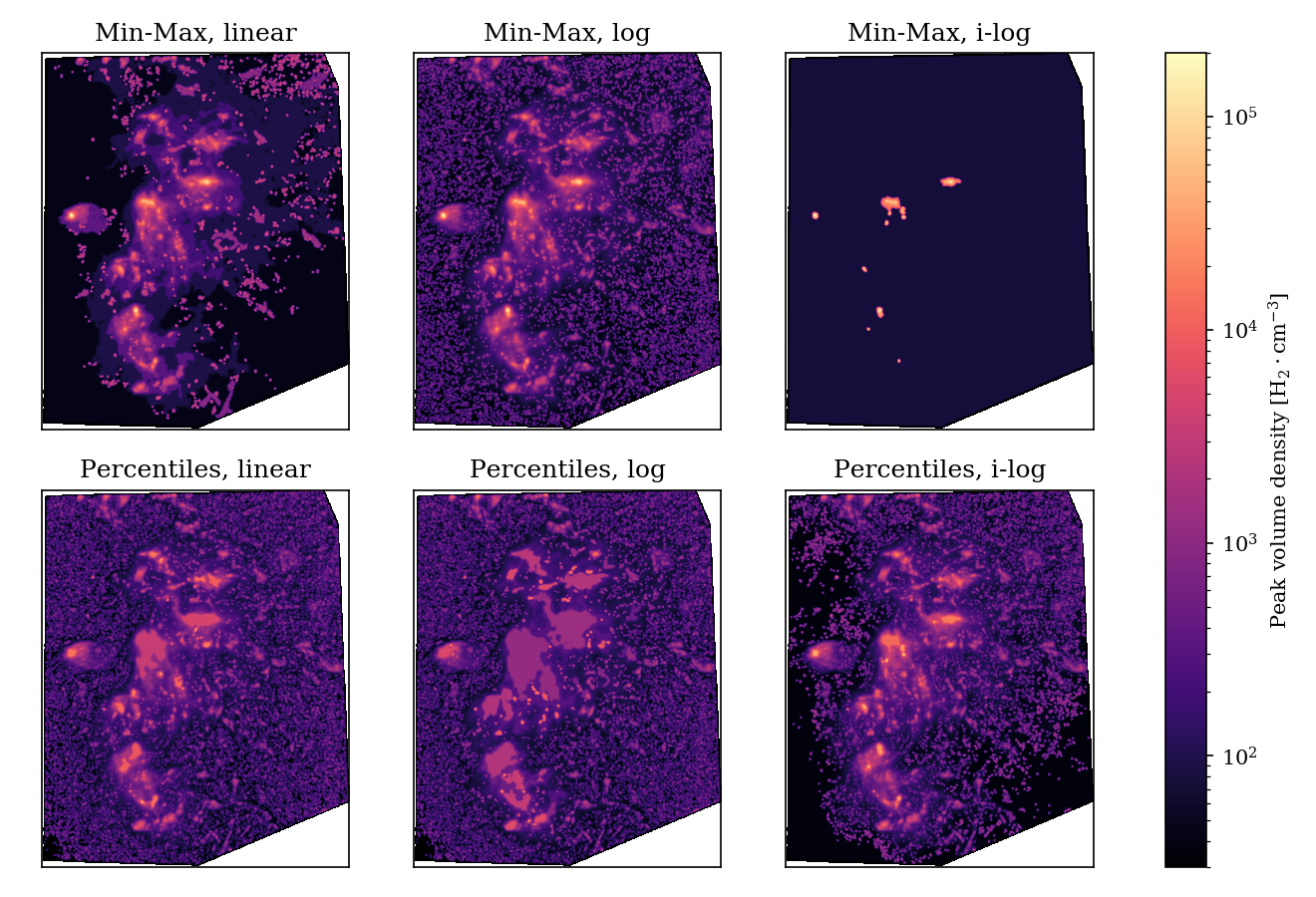}
        \includegraphics[width=\linewidth]{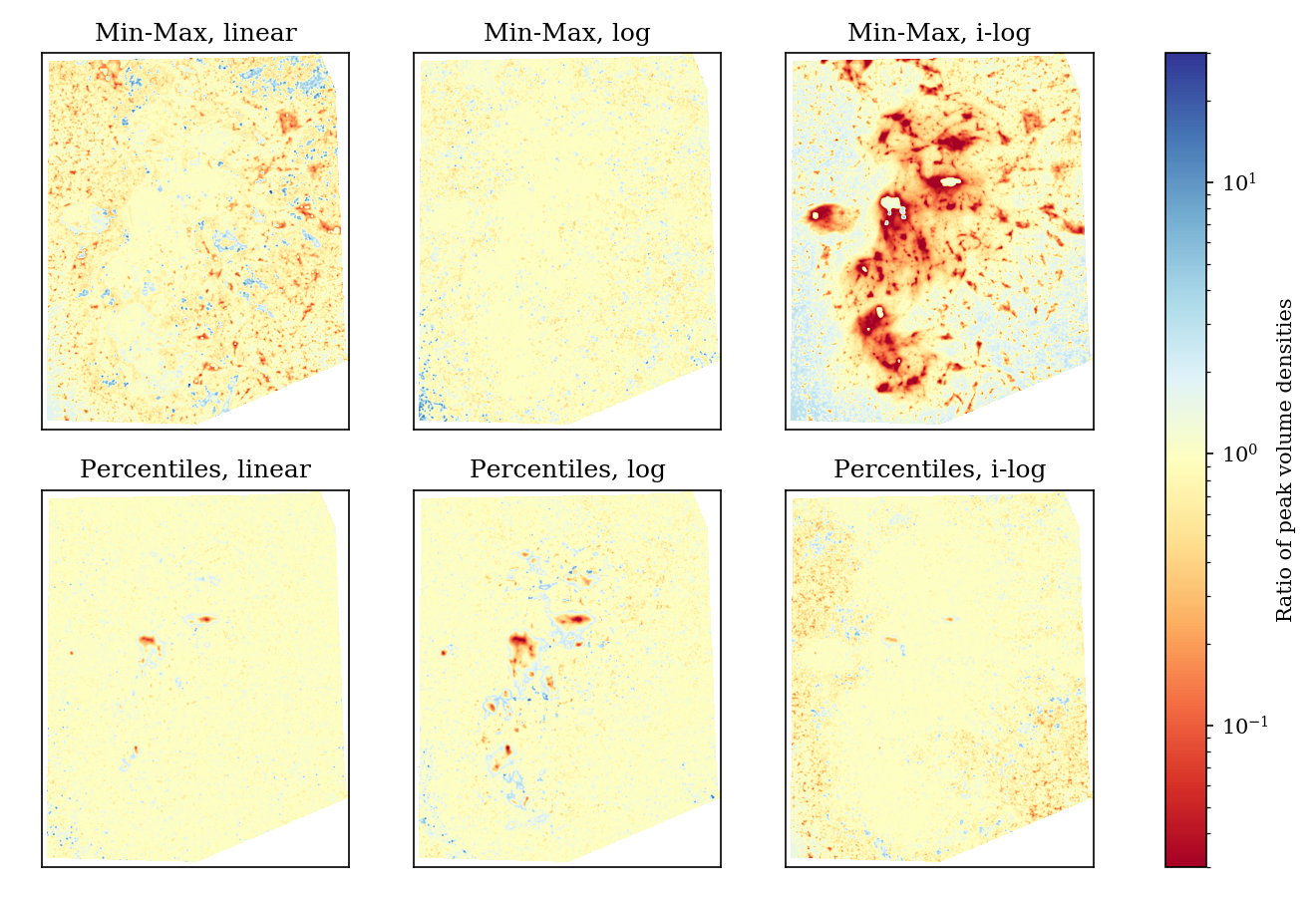}
        \caption{Effects of sampling options on the peak volume density maps, for the Cha II cloud. \emph{Top:} Peak volume density maps obtained with different coarse, 50-level volume density reconstructions. \emph{Bottom:} Ratio of these maps to the reference peak volume density map (obtained with a 2000-level i-log percentile sampling).}
        \label{fig:app:samplingmap}
    \end{figure}
}

\newcommand{\FigAppAlpha}{
\begin{figure}
    \centering
    \includegraphics[width=0.9\linewidth]{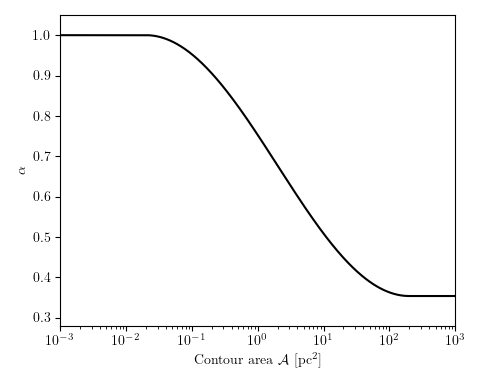}
    \caption{Variation of anisotropy factor $\alpha$ as a function of scale in the case of Orion\,A.}
    \label{fig:app:alpha}
\end{figure}
}

\newcommand{\FigAppRescMaps}{
\begin{figure*}
    \centering
    \includegraphics[width=\linewidth]{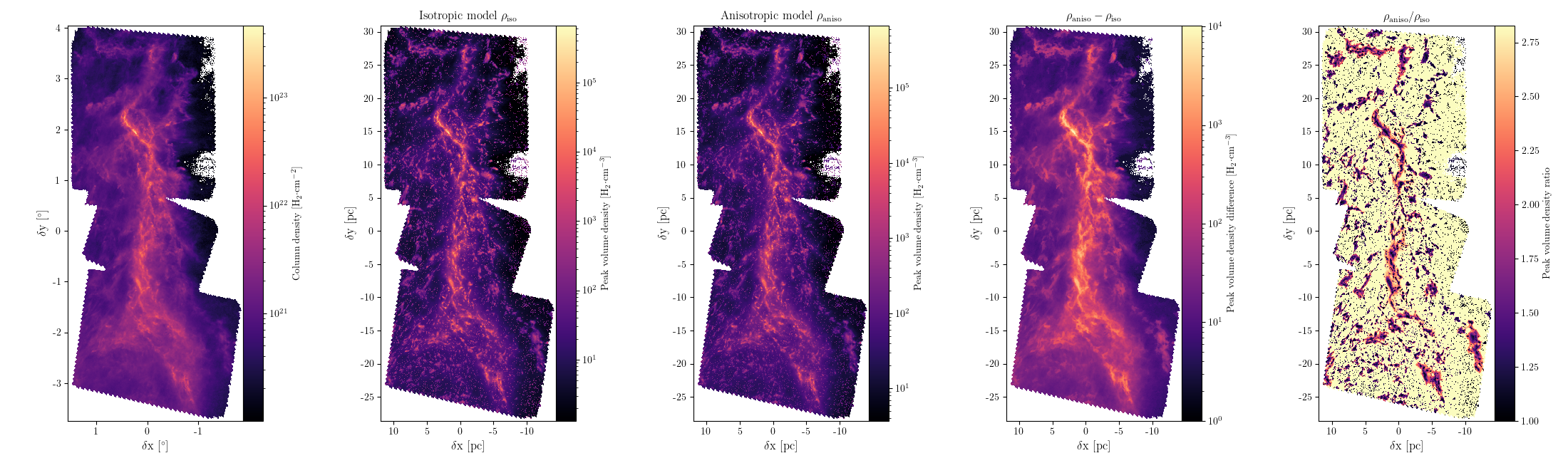}
    \caption{Comparison of the effect of scale-dependent anisotropy on density maps. \emph{From left to right:} Column density map used as a prior; peak volume density map obtained with the isotropic model; peak volume density map obtained with the anisotropic model; difference between the two peak volume density maps; ratio between the two peak volume density maps.}
    \label{fig:app:rescmaps}
\end{figure*}
}

\newcommand{\FigAppRescHist}{
\begin{figure}
    \centering
    \includegraphics[width=\linewidth]{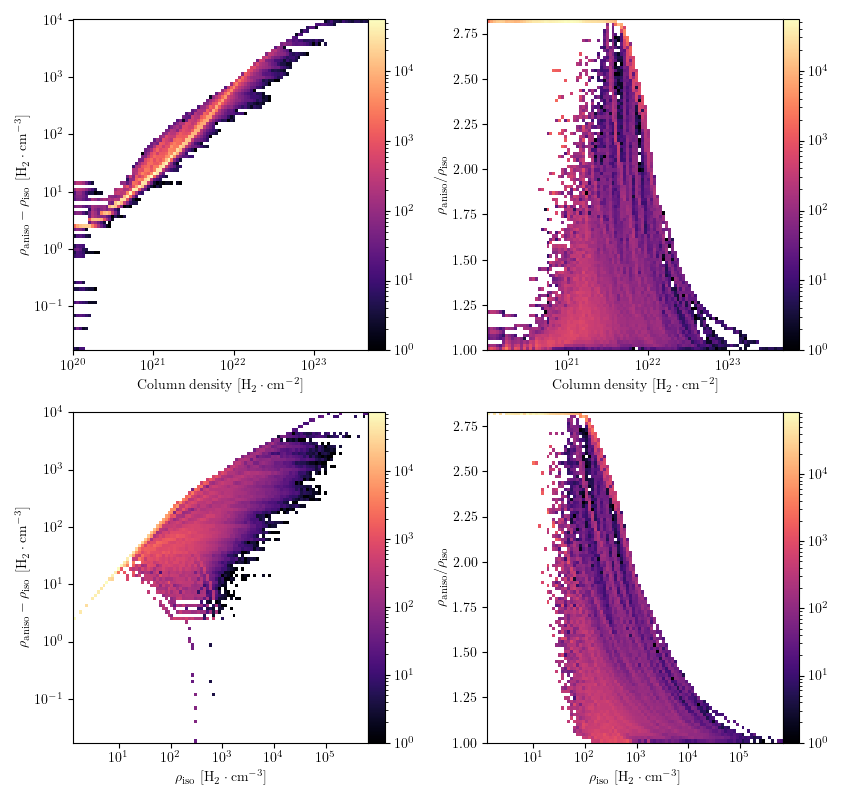}
    \caption{Joint distribution of the column density (\emph{top}) or isotropic peak volume density (\emph{bottom}) with the difference between the isotropic and anisotropic peak volume densities (\emph{left}) or their ratio (\emph{right}).}
    \label{fig:app:reschist}
\end{figure}
}

\newcommand{\FigAppRescPDF}{
\begin{figure}
    \centering
    \includegraphics[width=\linewidth]{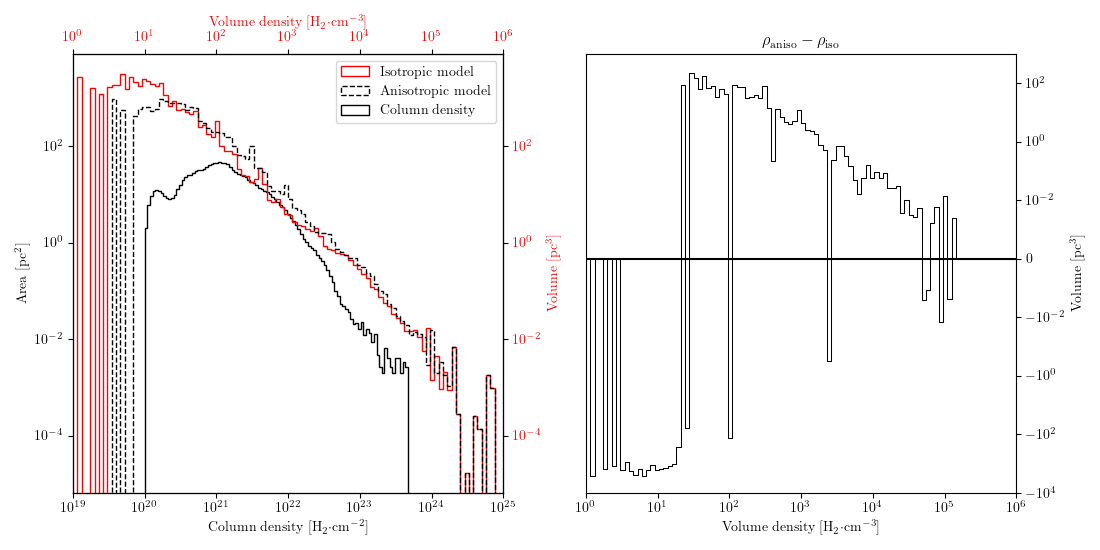}
    \caption{\emph{Left:} Comparison of the column density PDF with the global isotropic and anisotropic volume density PDFs derived for Orion\,A. \emph{Right:} PDF of the difference between the global volume density distribution derived by the anisotropic and isotropic model (presented on a symmetric logarithmic scale).}
    \label{fig:app:rescpdf}
\end{figure}
}

\newcommand{\FigAppOriA}{
\begin{figure*}
    \centering
    \includegraphics[width=0.48\linewidth]{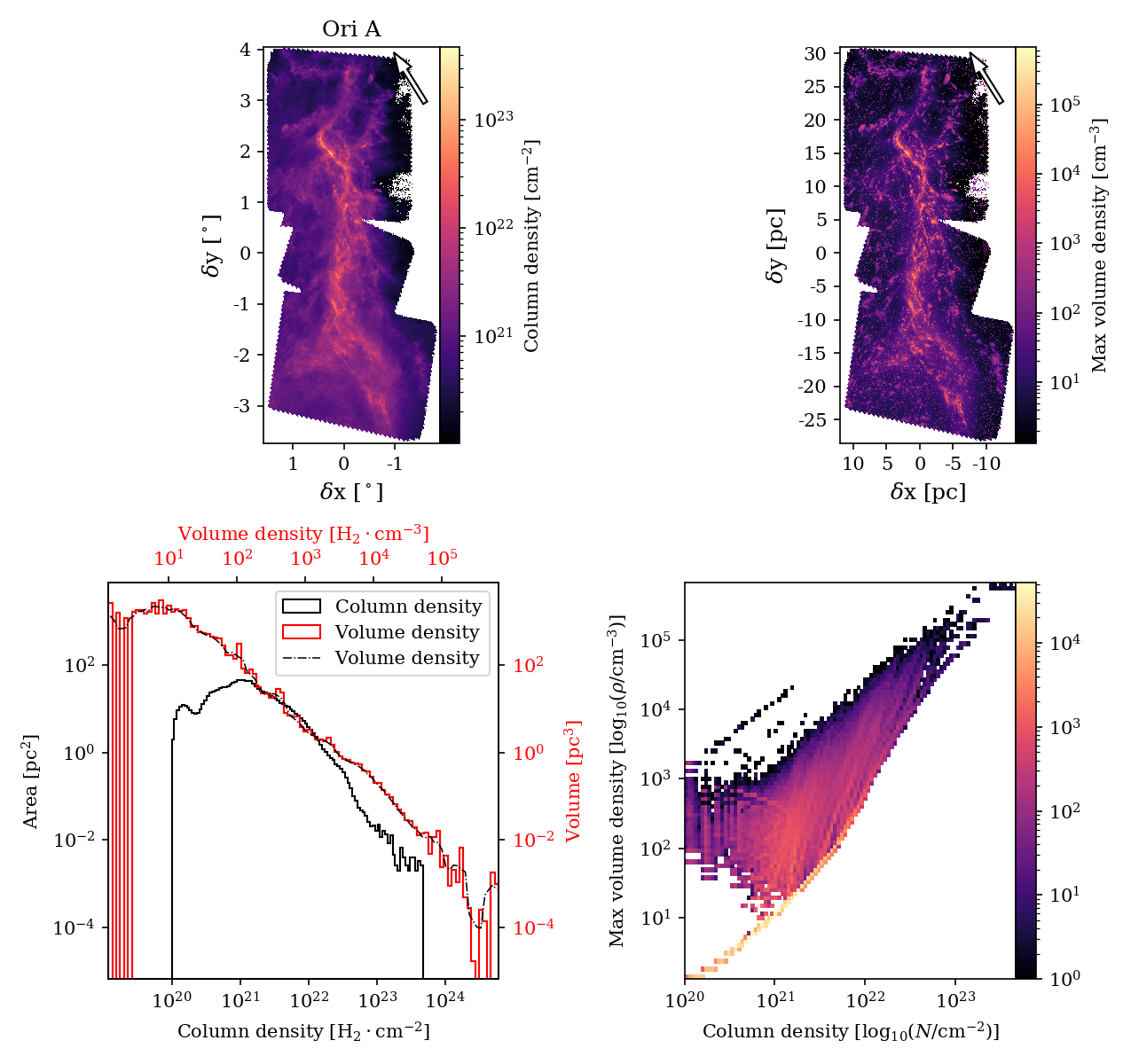}
    \includegraphics[width=0.48\linewidth]{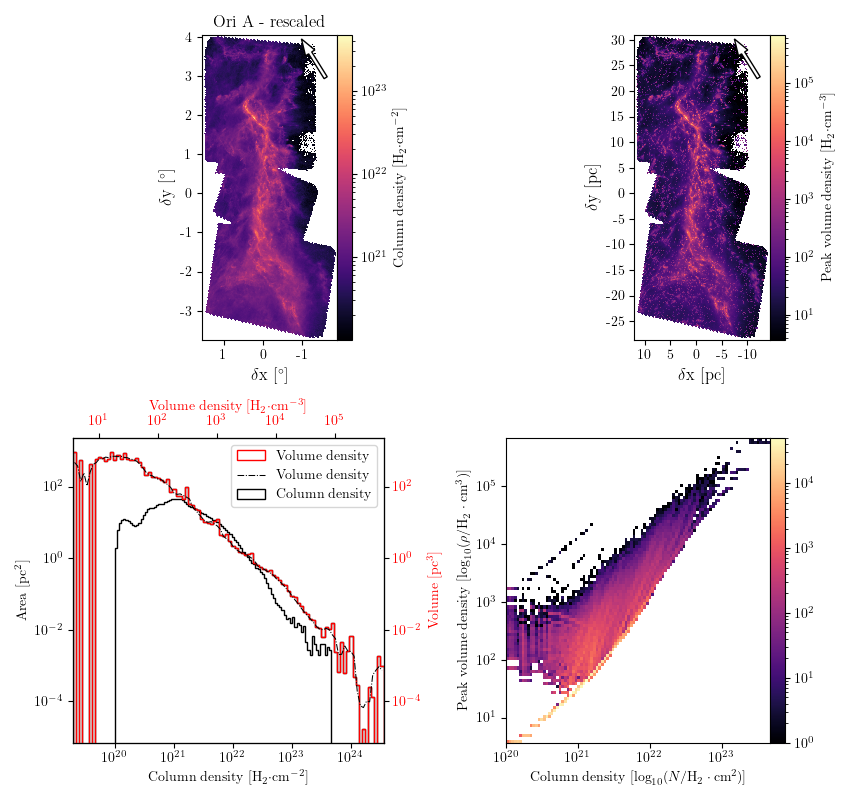}
    \caption{\emph{Left:} Same four panels as Fig. \ref{fig:panels}, for the Orion A giant molecular cloud. \emph{Right:} Same, but with anisotropy taken into account.}
    \label{fig:app:oria}
\end{figure*}
}


\newcommand{\clouddata}{
  \begin{table*}
    \centering
    \caption{Cloud sample and basic properties collected or directly derived from the literature.}
    \begin{tabular}{cccccc}
      \hline
      \hline
	  Cloud & Distance\tablefootmark{a} & Mass\tablefootmark{b} & YSO & SFR & SFE\\
	  \hline
	   & pc & $10^3$\Msun & \# & \Msun/\Myr & \\
	  \hline
 Aql & 487 & 89.162 & 416 \tablefootmark{d} & 104.0 & 0.23\% \\
 B68 (\object{LDN 57}) & 125\tablefootmark{b}  & 0.016 & 0 \tablefootmark{e} & 0.0 & 0.00\% \\
 Cep1157 & 341 & 0.633 & 6 \tablefootmark{d} & 1.5 & 0.47\% \\
 Cep1172 & 359 & 0.838 & 35 \tablefootmark{d} & 8.75 & 2.05\% \\
 Cep1228 & 364 & 0.961 & 14 \tablefootmark{d} & 3.5 & 0.72\% \\
 Cep1241 & 344 & 1.667 & 0 \tablefootmark{d} & 0.0 & 0.00\% \\
 Cep1251 & 336 & 0.827 & 32 \tablefootmark{d} & 8.0 & 1.90\% \\
 Cha I & 210 & 2.288 & 71 \tablefootmark{d} & 17.75 & 1.53\% \\
 Cha II & 190 & 1.197 & 21 \tablefootmark{d} & 5.25 & 0.87\% \\
 Cha III & 161 & 0.722 & 1 \tablefootmark{d} & 0.25 & 0.07\% \\
 CrA & 155 & 0.408 & 37 \tablefootmark{d} & 9.25 & 4.34\% \\
 IC5146 & 730 & 5.816 & 117 \tablefootmark{d} & 29.25 & 1.00\% \\
 L1689B & 154 & 0.139 & 1 \tablefootmark{d} & 0.25 & 0.36\% \\
 Lup I & 151 & 0.304 & 11 \tablefootmark{d} & 2.75 & 1.78\% \\
 Lup III & 197 & 0.333 & 43 \tablefootmark{d} & 10.75 & 6.06\% \\
 Lup IV & 108 & 0.130 & 5 \tablefootmark{d} & 1.25 & 1.88\% \\
 Mus & 190 & 0.695 & 2 \tablefootmark{d} & 0.5 & 0.14\% \\
 Oph L1688 & 139 & 3.853 & 238 \tablefootmark{d} & 59.5 & 3.00\% \\
 \object{Ori A} & 438 & 40.072 & 2640.3 \tablefootmark{f} & 660.08 & 3.19\% \\
 \object{Ori B} & 436 & 24.016 & 530.1 \tablefootmark{f} & 132.51 & 1.09\% \\
 Per & 294 & 7.990 & 345 \tablefootmark{d} & 86.25 & 2.11\% \\
 Pipe & 180 & 2.881 & 17 \tablefootmark{e} & 4.25 & 0.29\% \\
 Polaris & 343 & 0.206 & 0 \tablefootmark{g} & 0.0 & 0.00\% \\
 Ser & 487 & 60.450 & 236 \tablefootmark{d} & 59.0 & 0.19\% \\
 Tau L1495 & 130 & 1.948 & 84 \tablefootmark{h} & 21.0 & 2.11\% \\
	  \hline
	\end{tabular}
	\tablefoot{
	\tablefoottext{a}{Zucker et al. (2020) except where noted, see text for details;} \tablefoottext{b}{de Geus et al. (1989);} \tablefoottext{c}{integrated from HGBS column densities;} \tablefoottext{d}{Dunham et al. 2015;} \tablefoottext{e}{Forbrich et al. 2009;} \tablefoottext{f}{Megeath et al. 2012, see text for details;} \tablefoottext{g}{Ward-Thompson et al. 2010, André et al. 2010;} \tablefoottext{h}{Rebull et al. 2010}
    }
    \label{tab:clouddata}
  \end{table*}}

\newcommand{\TabColVol}{
    \begin{table}
        \caption{Properties of the correlation between column density and peak volume density in the Gould Belt for different density ranges.}
        \centering
        \begin{tabular}{ccc}
            \hline
            \hline
            Column density range & Power-law exponent & RMS scatter \\
            \hline
            $10^{21}$ H$_2\pscm$ & & dex \\
            \hline
            $0.1-500$ & $1.084 \pm 0.0004$ & 0.296\\
            $1.11-8.84$ & $0.918 \pm 0.001$ & 0.305\\
            $> 8.84$ & $2.218 \pm 0.007$ & 0.156
        \end{tabular}
        \label{tab:colvol}
    \end{table}

}

\newcommand{\TabDense}{
    \begin{table}
        \centering
        \caption{Dense gas measurements for all the clouds in the sample.}
        \begin{tabular}{ccccc}
            \hline
            \hline
            Cloud & $f_\emr{dg}^N$ & $f_\emr{dg}^\rho$ & $\Delta N$ & $\Delta \rho$ \\
            \hline
             & \% & \% & & \\
            \hline
            Aql & 11.77 & 0.36 & 0.57 & 1.29 \\
            B68 & 1.14 & 2.72 & 0.71 & 1.39 \\
            Cep1157 & 4.85 & 0.66 & 1.06 & 2.29 \\
            Cep1172 & 4.30 & 0.71 & 0.91 & 1.86 \\
            Cep1228 & 6.11 & 0.44 & 1.14 & 2.10 \\
            Cep1241 & 0.00 & 0.00 & 0.46 & 1.10 \\
            Cep1251 & 20.36 & 1.73 & 1.60 & 2.76 \\
            Cha I & 7.17 & 0.59 & 0.94 & 1.76 \\
            Cha II & 6.11 & 1.15 & 0.95 & 1.95 \\
            Cha III & 0.75 & 0.27 & 0.74 & 1.67 \\
            CrA & 20.37 & 3.51 & 1.53 & 2.74 \\
            IC5146 & 9.09 & 0.55 & 1.10 & 2.41 \\
            L1689B & 6.50 & 1.55 & 0.56 & 1.06 \\
            Lup I & 5.13 & 1.60 & 1.21 & 2.34 \\
            Lup III & 5.22 & 1.11 & 0.93 & 1.69 \\
            Lup IV & 4.65 & 1.48 & 0.95 & 1.97 \\
            Mus & 0.35 & 0.05 & 0.67 & 1.39 \\
            Oph L1688 & 12.00 & 2.25 & 1.08 & 2.05 \\
            Ori A & 24.90 & 3.40 & 1.47 & 3.10 \\
            Ori B & 13.15 & 2.23 & 1.26 & 2.64 \\
            Per & 13.36 & 2.90 & 1.32 & 2.72 \\
            Pipe & 0.81 & 0.29 & 0.38 & 1.12 \\
            Polaris & 0.00 & 0.00 & 0.46 & 1.26 \\
            Ser & 3.77 & 0.53 & 0.58 & 1.44 \\
            Tau L1495 & 5.60 & 0.88 & 0.94 & 1.97 \\
        \end{tabular}
        \label{tab:dense}
    \end{table}
}

\newcommand{\TabSFE}{
    \begin{table}
        \centering
        \caption{Fitting results of the correlation between SFE and the different dense gas descriptors.}
        \begin{tabular}{cccc}
            \hline
            \hline
            Quantity & Slope & Intercept & RMS scatter \\
            \hline
            $f_\emr{dg}^N$ & $0.75 \pm 0.18$ & $-0.59 \pm 0.14$ & 0.236 dex\\
            $f_\emr{dg}^\rho$ & $0.89 \pm 0.17$ & $0.03 \pm 0.006$ & 0.197 dex\\
            $\Delta N$ & $1.09 \pm 0.26$ & $-1.10 \pm 0.26$ & 0.189 dex \\
            $\Delta \rho$ & $0.56 \pm 0.16$ & $-1.15 \pm 0.33$ & 0.267 dex
        \end{tabular}
        \label{tab:sfe}
    \end{table}

}


\title{On volume density and star formation in nearby molecular clouds}

\author{
    Jan H. Orkisz \inst{\ref{IRAM},\ref{Chalmers}}{\thanks{orkisz@iram.fr}}
    \and
    Jouni Kainulainen \inst{\ref{Chalmers}}
}

\institute{%
  IRAM, 300 rue de la Piscine, 38 406 Saint-Martin-d’Hères, France \label{IRAM}
  \and Chalmers University of Technology, Department of Space, Earth and Environment, SE-412 93 Göteborg, Sweden \label{Chalmers} 
}

\date{Submitted 31 December 2022 / Accepted 3 December 2024}

\abstract%
{Volume density is a key physical quantity controlling the evolution of the interstellar medium (ISM) and star formation, but it cannot be accessed directly by observations of molecular clouds.} %
{We aim at estimating the volume density distribution in nearby molecular clouds, to measure the relation between column and volume densities and to determine their roles as predictors of star formation.} %
{We develop an inverse modelling method to estimate the volume density distributions of molecular clouds. We apply this method to 24 nearby molecular clouds for which column densities have been derived using \emph{Herschel} observations and for which star formation efficiencies (SFE) have been derived using observations with the \emph{Spitzer} space telescope. We then compare the relationships of several column- and volume-density based descriptors of dense gas with the SFE of the clouds.} %
{We derive volume density distributions for 24 nearby molecular clouds, which represents the most complete sample of such distributions to date. The relationship between column densities and peak volume densities in these clouds is a piece-wise power-law relation that changes its slope at a column density of $5 - 10 \e{22} \Htpccm$. We interpret this as a signature of hierarchical fragmentation in the dense ISM. We find that the volume-density based dense gas fraction is the best predictor of star formation in the clouds, and in particular, it is as anticipated a better predictor than the column-density based dense gas fraction. We also derive a volume density threshold density for star formation of $2\e 4$ H$_2\pccm$.} %
{}


\maketitle{}

\section{Introduction}

Volume density is one of the most important parameters affecting physical processes in the interstellar medium (ISM), from chemistry to thermodynamics, particle interactions, radiation mechanisms, and gas dynamics. Volume density is therefore a crucial parameter in any evolution model of the ISM, and consequently, for any model of star formation.

Unfortunately, it is not possible to directly measure volume densities at the densities of typical star formation regions within molecular clouds ($n(\mathrm{H}_2) \gtrsim 10^2$ cm$^{-3}$). Novel techniques exploiting Gaia data can infer the 3-dimensional structure of the ISM using dust, but only at relatively low spatial resolutions ($\gtrsim 1\pc$) and volume densities ($\lesssim 1 \Htpccm$) that do not reach the actual birthplaces of stars \citep[\eg][]{green19, leike20, rezaei22, lallement22}. In contrast, at the densities intimately linked to star formation, we only have access to measurements of column densities. As a result, the detailed volume density structure of star-forming regions and its properties remain poorly known. This, in turn, hampers our understanding of the physics of ISM evolution and star formation. 

When the geometry of the studied region can be assumed to be simple, such as a cylinder or a sphere, determination of the volume density distribution can be done trivially via forward modelling \citep[\eg][]{alves01,kainulainen16,orkisz19,suri19,arzoumanian19, konyves20}. However such assumptions are not generally applicable to the complex density structures of molecular clouds. Similar approaches can be applied to galactic scales via assumptions about the gas geometry in the disk; indeed a volumetric approach of star formation laws has proved to yield more robust correlation between gas and stars than the usual, surface-density based approach \citep[\eg][]{bacchini19a,bacchini19b,yim22}.

A number of  more or less sophisticated inverse modelling approaches, that is methods which use the observed column densities to construct a model of the underlying volume densities, have been presented \citep[\eg][]{kainulainen14,krco16,bron18,hasenberger20}. One should also note the work of \citet{xu23}, who combine the use of magneto-hydrodynamic simulations and deep-learning techniques to predict mean volume densities of clouds based on the column density maps. However overall, the development of such methods is clearly still in progress and none of the existing techniques has been systematically applied to major sets of observational data.

In this paper, we will make progress by developing a method, \coltt{} (``from COLUMn to vOLUME''), to estimate volume density structure of molecular clouds. We will then apply it to an extensive set of 24 molecular clouds observed with the \emph{Herschel} satellite. This enables us to present a new systematic study of the volume density statistics of molecular clouds, and specifically, the first one employing a complete set of the \emph{Herschel} column density data available for nearby clouds. With the resulting density data in hand, we will address whether star formation in the clouds is better traced by volume or column densities, and we also derive a volume density threshold for star formation.

The paper is organised as follows. In Section \ref{sec:data-method}, we present in details the principles and implementation of the \coltt{} code, as well as the molecular cloud data to which we applied \coltt{} and the star formation data that we confronted with our volume density estimations. Section \ref{sec:results} presents the obtained volume density estimations, as well as the observed correlations between column density, volume density and star formation. The implications of these findings, as well as the possible caveats, are discussed in Section \ref{sec:discussion}. Our findings are then summarised in Section \ref{sec:conclusion}. This paper also has an Appendix \ref{app:simulations} which describes how we tested the \coltt{} code on simulation data, an Appendix \ref{app:sampling} discussing technical details of the implementation of \coltt{}, and an Appendix \ref{app:aniso} discussing the impact of cloud anisotropy on the results.

\section{Data and methods} 
\label{sec:data-method}
\subsection{Volume density estimation} 
\label{sec:colume}

In this section we describe the principle of \coltt{}, the volume density estimation algorithm that we implemented and applied to column density data of nearby molecular clouds. The code will be made publicly available. It has been implemented for the most part using \texttt{NumPy} \citep{numpy} in Python3. The \coltt{} code makes also extensive use of the \texttt{scikit-image} package \citep{scikit-image} for image analysis (basic image processing, region identification and measurements), as well as the \texttt{AstroPy} package \citep{astropy13,astropy18} for astronomical data handling.

\subsubsection{Basic principle and geometrical assumptions} 
\label{sec:geometry}

Astronomical observations only give us access to the column density $N$ (of gas, dust, etc.) which, in the simplest case, is a projection of the volume density field $\rho$ in the plane of the sky:
\begin{equation}
    N(x,y) = \int\!\rho(x,y,z)\diff z
\end{equation}
where $z$ is along the line of sight. In the lack of knowledge about the three-dimensional structure of the volume density field subtending the column density map available to the observer, we have to make assumptions about what is deemed to be a probable geometry. In the case of the \coltt{} code, the assumptions are as simple as possible and are the following:
\begin{itemize}
    \item The volume density structure is single-peaked along each line of sight.
\end{itemize}
This is the simplest and safest assumption when one cannot tell if a cloud has several components along the line of sight: even with access to velocity data, it has been shown that the conversion from position-position-velocity (PPV) to position-position-position (PPP) is not reliable \citep[\eg][]{clarke18}. This first assumption entails than the low density gas forms the fore- and background of the high density gas, at all densities and for all lines of sight, and thus:
\begin{itemize}
    \item The volume density structure is hierarchical.
\end{itemize}
This intuitively means that very high density regions are always nested within regions of high density, themselves embedded in regions of intermediate density, etc.

Last but not least, it is necessary to estimate the dimension of volume density structures along the line of sight. The simplest and most common assumption here is that it is typically the same as the dimensions in the plane of the sky, and thus:
\begin{itemize}
    \item The volume density distribution is statistically isotropic.
\end{itemize}

The statistical isotropy implies that the dimension of a object along the line of sight (the depth) is comparable to its dimensions as seen in the plane of the sky. There as several ways to implement this requirement, and we adopted in \coltt{} what is perhaps the simplest one:
\begin{equation}
    \mathcal{V} = \alpha \mathcal{A}^{3/2} \Leftrightarrow l = \alpha \sqrt{\mathcal{A}}
\end{equation}
where $\mathcal{A}$ is the area of an object in the plan of the sky, $l$ is its depth along the line of sight and $\mathcal{V}$ is its total volume. The factor $\alpha$, of the order of 1, depends on the detailed geometry -- for the sake of simplicity we have set $\alpha = 1$, which strictly speaking is only valid for a face-on cube geometry (other options for $\alpha$ are discussed in Appendix \ref{app:aniso}).
\\

These assumptions allow us to work with an incremental approach, where instead of considering the column density $N$ and the volume density $\rho$ directly, we rather describe them as sums of infinitesimal (or, in practice, finite) increments of density. In that way, we can write for any line of sight $(x,y)$ that:
\begin{equation}
    N(x,y) = \sum_{i=0}^{i_\mathrm{max}(x,y)}\!\Delta N_i = \sum_i N_{i+1} - N_i
\end{equation}
The sampling $\Delta N_i$ does not need actually to be regular (see Sect. \ref{sec:implementation} and Appendix\ref{app:sampling} for a discussion).

Likewise, the volume density in all points of space would be defined as:
\begin{equation}
    \rho(x,y,z) = \sum_{i=0}^{i_\mathrm{max}(x,y)}\!\Delta \rho_i(x,y,z)
\end{equation}

It is therefore these sums of $\Delta\rho_i(x,y,z)$ that we aim to reconstruct. However, let us stress that we do not claim to recover actual density profiles along the $z$ direction, we only try to infer the statistical distribution of volume density for each line of sight.

In order to reconstruct the volume density increments, we consider the contour $\mathcal{C}_i$ associated with each column density increment $\Delta N_i$. Each such contour is considered as an individual, isotropic object, of area $\mathcal{A}_i$, embedded in the contour $\mathcal{C}_{i-1}$ and itself providing a background for the contour $\mathcal{C}_{i+1}$. In three dimensions, this corresponds to volume density increments with associated volumes hierarchically nested within one another. The volume density increments $\Delta\rho$ are therefore defined as:
\begin{equation}
\label{eq:Drho}
    \Delta\rho_i = \frac{\Delta N_i}{\sqrt{\mathcal{A}_i}}
\end{equation}

Equation \ref{eq:Drho} implements the assumptions of our method, by setting the depth aspect ratio to $\alpha=1$, and ensuring that the volume density increment is not fragmented along the line of sight. Each volume density increment $\Delta\rho_i$ is assumed then assumed to be fully embedded in the volume of the increment $\Delta\rho_{i-1}$ (where $N_i > N_{i-1}$). This excludes for example the possibility of an isolated dense clump in the diffuse foreground of a cloud.
\\

An additional complexity comes from the fact that in practice the contours are not necessarily connected, but instead consist of several disconnected regions which need to be treated individually. Taking this into account makes the volume density estimation more computationally intensive, as it requires a topological analysis of each contour, but the cost is not prohibitive. The principle and assumptions of the computation remain exactly the same as described above, with the only difference that each sub-region $\mathcal{R}_{i,n}$ identified in a fragmented contour is treated in the way the entire $\mathcal{C}_i$ was treated before. This has the effect of giving to Eq. \ref{eq:Drho} an additional spatial dependence beside the value of $N(x,y)$. Each sub-region $\mathcal{R}_{i,n}$ now has an corresponding area $\mathcal{A}_{i,n}$, which implies:

\begin{equation}
    \Delta\rho_i(x,y) = \frac{\Delta N_i}{\sqrt{\mathcal{A}}_{i,n}(x,y)}
\end{equation}
where $\mathcal{A}_{i,n}(x,y) = \mathcal{A}_{i,n}$ such that $(x,y) \in \mathcal{R}_{i,n}$.

Note that here, despite taking into account fragmentation of column density contours in the plane of the sky, we still assume that the identified regions correspond to connected volumes, meaning that there is no fragmentation along the line of sight.
\\

From these, one can in particular compute the maximum volume density reached along each line of sight (``peak volume density'' hereafter):
\begin{equation}
\label{eq:rhopeak}
    \rho_\mathrm{peak}(x,y) = \sum_{i=0}^{i_\mathrm{max}(x,y)} \! \frac{\Delta N_i}{\sqrt{\mathcal{A}_{i,n}(x,y)}}\,, \; \mathrm{where} \; N_{i_\mathrm{max}(x,y)} = N(x,y)
\end{equation}

The distribution (probability distribution function, PDF) of volume densities along each line of sight can also be described by the list of volume density values $\rho_i(x,y)$, weighted by the corresponding depth along the line of sight $l_i(x,y)$. The depth $l_i(x,y)$ describes how much space along the line of sight is found at a volume density of $\rho_i(x,y)$. For any value of $\rho_i \leq \rho_\mathrm{peak}(x,y)$, the quantities are simply obtained as:
\begin{equation}
\label{eq:rho}
\rho_i(x,y) = \sum_{k=0}^i \Delta \rho_k(x,y)
\end{equation}
\begin{equation}
\label{eq:lrho}
l_i(x,y) = \sqrt{\mathcal{A}_{i+1,n}(x,y)} - \sqrt{\mathcal{A}_{i,n}(x,y)}
\end{equation}

\FigToyGeometry{}

Figure \ref{fig:toygeometry} illustrates the geometrical principle of the volume density estimation for a toy example in the disconnected case (taking into account plane-of-the-sky fragmentation). The column density map (here containing only three discrete levels) is decomposed into column density increments, and the area of the contours or sub-regions thereof is measured. The column density increments are then converted into volume density increments by attributing them with a depth along the line of sight, and the final density structure is reconstructed by summing the volume density increments. The resulting volume density estimation is represented in Fig. \ref{fig:toygeometry} as a three-dimensional rendering for illustration purposes only -- the produced volume density datasets are instead maps of the volume density PDF for each line of sight. Such PDFs are shown for several lines of sight of our toy example in Fig. \ref{fig:toypdf}. In practice the depth $l_i(x,y)$ can be converted to a volume $v_i(x,y)$ by multiplying it by the physical area of a resolution element (pixel) on the sky.

\FigToyPDF{}

\subsubsection{Practical implementation aspects}
\label{sec:implementation}

Defining column density increments requires a finite number of column density levels. The highest possible number of levels is dictated by the number of individual values found in the input data. With noisy data and a high-resolution quantisation, this can be as many as the number of pixels in the map -- in such a case each contour is smaller than the previous one by only one pixel. The way volume densities are estimated by the \coltt{} code require to have in memory the maps corresponding to each contour at the same time, which means $X\times Y\times S$ data-points, where $X\times Y$ are the native dimensions of the column density map, and $S$ is the column density sampling, corresponding to the number of column density contours used. With the native (\ie{} maximal) number of levels, this means effectively close to $\left(X \times Y\right)^2$ data-points, which leads to unmanageable computations in terms of random-access memory requirements, even for fields of moderate sizes. It is therefore crucial to reduce the data complexity by sampling the column density distribution in an efficient way, that is in a way which recovers as much as possible of the detailed cloud structure while using as few column density levels as possible.

The choice of the sampling method depends among other things on the range of the column density distribution that need to be described most accurately. In this work, we have used the same sampling for all the studied column density maps. The sampling method is based on the column density percentiles (which ensures that there are no empty bins), with 2000 bins and an inverse logarithmic (i-log) spacing. The i-log spacing is the logarithm of a linear spacing, scaled to the range of the data (or, in that case, from 0 to 100 as we are considering percentiles). This sampling method and other possible sampling choices are discussed in Appendix \ref{app:sampling}. At this stage the data is compressed by posterisation: in each bin, the individual column density values are replaced by the average value within the bin. In the column density map, this effectively creates contours, which are then used to compute areas and volumes as described in Sect. \ref{sec:geometry}.\\

Once all the depths along the line of sight and all volume density increments have been computed, the yielded data structure is made of two three-dimensional arrays: a map of the lists of volume densities for each line of sight, and a map of the lists of associated volumes (physical area of a pixel $\times$ depth along the line of sight) -- which can together be described as a map of weighted volume density PDFs as described in Eq. \ref{eq:lrho}. The dimension of these arrays is $X\times Y \times S$. If the column density is very finely sampled, the resulting arrays are therefore not only very large, but also very sparse, since only the one point with the highest column density value is enclosed in all the contours and is therefore described to the highest complexity levels -- each line of sight is described by as only many volume density levels as the number of contours it was enclosed in. 
A compromise between saving two impractically large file or an impractically high number of small files (two per line of sight) is to save the computation results on a row-per-row basis in individual, two-dimensional \texttt{NumPy}-array files. A folder with $2X+1$ files is therefore generated -- the additional file being a copy of the header of the input column density map.

In addition to the comprehensive data structure described above, there is the possibility to extract only the basic information about the volume density in the cloud, namely the peak volume density, and the global volume density PDF (\ie{} the sum of the PDFs of all the lines of sight). This data format is a highly lossy compression, but it is also much lighter and easier to handle, as it only produces a two-dimensional map and a one-dimensional PDF, instead of two large three-dimensional data-cubes. It also offers the advantage of portability, since the map of peak volume densities can be saved in the FITS format, and the global PDF as a text file. Incidentally, this is also the data format used for the present study. This data format is illustrated in Fig. \ref{fig:toypanels}.\\

\FigToyPanels{}

Regarding the different geometrical treatments described in Sect. \ref{sec:geometry},  both the connected and disconnected treatment of contours are implemented in \coltt{}. While the disconnected estimator offers a more detailed treatment of the column density morphology, and is physically more realistic, the connected estimator has the interesting property of being a ``lower limit'' estimator for the peak volume densities. This is due to the fact that the column density increments are ascribed to the largest reasonably possible volume (\ie{} a connected, isotropic blob), whereas any modification (fragmentation, anisotropy) would tend to reduce this volume and therefore increase the volume density. Because all the lines of sight within a given column density contour are considered to belong to the same volume, it also produces a one-to-one correspondence between column and volume density values. In this work however, since we focus on star formation and therefore need accurate volume densities of dense gas in highly fragmented contours, we used exclusively the disconnected estimator.

\subsubsection{Method validation} 

In order to check whether the volume densities obtained with \coltt{} are reasonably close to the truth, we tested the reconstruction in a case were the 3D information is known, namely on numerical simulations. The tests were run on a $4\e3 \Msun{}$ cloud in a $40\times40\times40\pc$ cube, their results are described in detail in Appendix \ref{app:simulations}.

We have in particular checked the quality of the reconstructed peak and mean volume densities along the line of sight, and of the global volume density PDF, for projections along all three axes of the simulated cube, in both a noiseless and noisy case. The tests have in particular highlighted the reliability of the volume density PDF and the robustness of our isotropy assumption, since the reconstructed PDFs are very consistent from one projection to the other and show little sensitivity to noise.

While the average volume densities are very well reproduced, a systematic non-uniform bias is observed in the reconstruction of the peak volume densities. While the values are well reconstructed for low ($\sim 10^0 \pccm{}$) and high ($\gtrsim 10^4\pccm{}$) peak column densities, at intermediate values the peak volume density can be underestimated by as much as an order of magnitude. This is interpreted as being due to the fact that for the lowest and highest volume densities, we are dealing with the envelope of the cloud or dense cores respectively, which both comply well with the isotropy assumption, whereas between these cases the gas is structured into sheets and filaments which have high aspect ratios, their volumes thus get overestimated and densities consequently underestimated.

We also show that on average volume densities tend to get slightly biased towards higher values in the presence of noise, because the contours of the column density increments tend to get fragmented by the noise, which leads to smaller areas, thus smaller reconstructed volumes and higher densities. This effect is however moderate in comparison with the other uncertainties of the method, and is largely negligible with high \SNR{} data.

\subsection{Molecular cloud sample} 
In order to show the potential of \coltt{} volume density estimations in the study of ISM and star formation with observational astronomical data, we applied it to a sample of molecular clouds for which homogeneous and high quality column density maps as well as young stellar object (YSO)catalogues are available. The chosen cloud sample and the corresponding column density maps and YSO catalogues are described in the following sections.

\subsubsection{Column density maps} 
\label{sec:coldens}
In order to benefit from the bet possible conditions to run \coltt{}, we needed a sample of column density maps with a high \SNR, a high spatial resolution, a homogeneous data reduction, and a wide diversity of environments. These requirements led us to choose the column density maps published for the set of 25 dense clouds observed by the \emph{Herschel} Gould Belt Survey \citep[HGBS,][]{andre10}. These nearby clouds have masses ranging from 16\Msun{} to 89\e3\Msun, with very different levels of star formation activity.

The studied clouds, as published by the HGBS consortium, are the following: 
the \object{Aquila Rift} (W40) cloud complex \citep{bontemps10,konyves15}, the \object{Cepheus Flare} clouds \citep{difrancesco20}, the \object{Chamaeleon} cloud complex \citep{winston12,kospal12,alves14}, the \object{Corona Australis cloud} \citep{siciliaaguilar14,bresnahan18}, \object{IC5146} \citep{arzoumanian11,arzoumanian19}, the \object{Lupus cloud} complex \citep{rygl13,benedettini18}, the \object{Musca} cloud \citep{cox16}, the Ophiuchus \object{L1688} and \object{L1689B} clouds \citep{roy14,arzoumanian19,ladjelate20}, the Orion\,A \citep{roy13, polychroni13} and Orion\,B \citep{schneider13,konyves20} giant molecular clouds, the \object{Perseus cloud} \citep{pezzuto12,pezzuto21,sadavoy12,sadavoy14}, the \object{Pipe nebula} and B68 (\object{LDN 57}) globule \citep{peretto12,roy14}, the \object{Polaris Flare} \citep{menshchikov10,wardthompson10,mivilledeschenes10}, the \object{Serpens} (Serpens Main + Aquila East) cloud complex \citep{fiorellino21}, and the \object{Taurus molecular cloud} \citep{palmeirim13,kirk13,marsh16}.

In summary, the column density maps were obtained from the observational data in the following way (see the above-cited HGBS papers for details). The \emph{Herschel} observations were carried out in parallel in all the bands of the SPIRE (Spectral and Photometric Imaging Receiver) and PACS (Photodetector Array Camera and Spectrometer) instruments. After data reduction, zero-level offsets were adjusted using \emph{Planck} data. The spectral energy distributions (SED) thus obtained were then fitted on a pixel-by-pixel with a modified black-body, with a dust opacity law of $\kappa_\lambda = 0.1 \times \left(\lambda/300\micron\right)^{\beta}\unit{cm^2\,g^{-1}}$ and a dust emissivity index $\beta=2$ \citep{hildebrand83}, which produces column density and effective dust temperature for the entirety of the observed maps. The produced maps are expressed in \Htpscm{}, taking only into account the hydrogen contribution to the column density, without helium and metals. 

The maps provided by the HGBS consortium\footnote{\url{http://www.herschel.fr/cea/gouldbelt/en/Phocea/Vie_des_labos/Ast/ast_visu.php?id_ast=66}} were reprocessed before the \texttt{Colume} processing. Given that the volume density computations are very memory-intensive, it was crucial to reduce the size of the files to a minimum. For this purpose, the maps were rotated depending on the shape of the observed field, in a way which allowed cropping as much blank area as possible. The maps were also resampled to a homogeneous 12'' pixel size - a sufficient sampling given the typical beam size of 36.3'', but which allows for much smaller files than the original 3'' sampling. To account for artefacts caused by the resampling as well as to remove noisy pixels at the edges of the observed field, the maps were submitted to 3 iterations of binary erosion (\ie{} their blanking mask was dilated 3 times), which corresponds to a full beam width. The column density values below 1\e{20}\Htpscm{} were then blanked out, since several maps featured spurious values below this threshold, and column densities below this value are anyway largely irrelevant for star formation on the studied scales. After this blanking step, the maps were eroded once more to smooth out the effect of the thresholding.

The reframing applied in several HGBS papers to remove the noisy edges of the maps was not implemented, since it was largely redundant with the filtering described above.

\subsubsection{Distances to the clouds} 
\label{sec:distances}
For each cloud, the distance was adopted based on the catalogue created by \citet{zucker20}, which provides distances to the first major extinction jump along the line of sight, which corresponds to the close end of the dense cloud. Since it often happens that many lines of sight of the \citet{zucker20} catalogue fall within a given \emph{Herschel} field, we adopted as our reference distance the line of sight closest to the centre of projection of each field. In the case of the Perseus cloud, the centre of projection happens to coincide with a ``hole'' in the cloud, where the estimated distance is the furthest of all available lines of sight, about 20\% further than the average -- we therefore used in that case the average distance of all the lines of sight corresponding to the field. The variation in distance from one line of sight to the other within a field is generally of the order of $\sim10$\% (when applicable), which in turns reflects as 10\% of uncertainty on the volume densities (via the uncertainty on the depth along the line of sight which is directly proportional to the distance).

The only exception to this distance procedure is the B68 globule, which, in the absence of foreground stars, lacks a relevant line of sight in \citet{zucker20} -- the line of sight closest to the centre of projection of the B68 field is the same as for the Pipe Nebula. The distance to B68 was thus set based on the most recent available distance measurement, from \citet{degeus89}.

\subsection{YSO samples and star formation properties} 
\label{sec:yso}
A homogeneous catalogue of YSOs was not available for the entire sample of clouds in our study. We therefore had to aggregate data from several YSO surveys. To select these surveys, rather than choosing the most recent (and arguably most complete) ones, we decided to focus on obtaining a catalogue which would be as homogeneous as possible in terms of sensitivity and methodology, and therefore completeness. This implied to have observations and data reduction carried out in similar ways, and having access to detailed YSO catalogues, rather than YSO counts, to enable cross-checking. 

For almost all the clouds in our sample, the YSO data was obtained from the catalogue created by \citet{dunham15}, which compiles a large YSO catalogue based on the legacy data of the c2d \citep{evans03} and Gould's Belt \citep{gutermuth09} surveys. We complemented this with several other catalogues: for the Orion clouds (Orion\,A and Orion\,B), we used the study of \citet{megeath12}; for the Pipe nebula and the B68 globule, we used the data of \citet{forbrich09}; for the Taurus cloud, we used results from \citet{rebull10}. Finally, not a single YSO is known to be present in the Polaris Flare \citep{wardthompson10,andre10}. 

The main properties of each of these YSO catalogues are nearly identical. All studies are based on multi-band observations made with the MIPS \citep[Multiband Imaging Photometer for \emph{Spitzer},][]{rieke04} and IRAC \citep[InfraRed Array Camera,][]{fazio04} instruments of the \emph{Spitzer} space observatory. The data reduction follows standard \emph{Spitzer} pipelines, and the extracted point sources are matched with the Two-Millimetre All-Sky Survey \citep[2MASS,][]{skrutskie06} catalogue. The identification of YSOs is then in all cases performed using a combination of colour-colour diagrams (or spectral indices) and colour-magnitude diagrams across the MIPS and IRAC bands, although the details of the choice of diagrams optimal for this identification varies from study to study. The main difference between the YSO samples is the quality assessment of the YSO candidates performed by \citet{rebull10}: after a first selection made using exclusively \emph{Spitzer} and 2MASS data, the reliability of the YSO candidates was examined using additional observational data from the Canada-France-Hawaii Telescope (CFHT) and Sloan Digital Sky Survey (SDSS) in the optical, as well as \emph{XMM-Newton} in the X-ray and ultraviolet. To a lesser extent, the approach taken in \citet{megeath12} is also different from the rest of the sample in the way that completeness and contamination of the YSO sample are tested statistically using artificial stars and comparison with reference fields. However, overall the discrepancies between the assembled YSO catalogues are minor, and for lack of a single, uniform YSO catalogue of all nearby molecular clouds, they still constitute for our sample of clouds the best possible database for the study of star formation properties in terms of self-consistency and data quality.

For each catalogue, we checked the positions of the YSOs against the spatial coverage of the corresponding column density maps. The YSOs were then selected according to their spectral type, namely class 0, class I, flat-spectrum and class II, either using directly the spectral classification provided in the catalogues \citep{forbrich09,megeath12,rebull10}, or inferring the class from the spectral measurements \citep{dunham15}. Only the robustly identified YSOs were retained, which implies removing from the sample sources identified as likely asymptotic giant branch (AGB) stars \citep{dunham15}, removing rejected candidates \citep{forbrich09}, keeping only candidates with the highest confidence levels \citep[``new member'' or ``probable new member'' only][]{rebull10}. In the case of Orion\,A and Orion\,B, \citet{megeath12} estimate a residual contamination of 6.1 per deg$^2$ from extragalactic sources, and that a further 13 sources are likely misidentified as YSOs, we therefore subtracted these contaminants from the total YSO count in each field by multiplying 6.1 by the field area and attributing the remaining 13 sources based on the pro-rata of YSOs identified in Orion\,A and Orion\,B -- this leads to non-integer YSO numbers in these two clouds.

The star formation rate (SFR) and star formation efficiency (SFE) of the clouds are then derived as
\begin{equation}
\label{eq:sfr-sfe}
    S\!F\!R = \frac{N_\emr{YSO}\cdot M_\emr{YSO}}{\tau_\emr{YSO}} \;\; \emr{and} \;\; S\!F\!E = \frac{N_\emr{YSO}\cdot M_\emr{YSO}}{N_\emr{YSO}\cdot M_\emr{YSO} + M_\emr{cloud}}
\end{equation}
where $N_\emr{YSO}$ is the YSO count for each cloud, $M_\emr{YSO} = 0.5 \Msun$ is the average YSO mass \citep{chabrier03}, $\tau_\emr{YSO} = 2\Myr$ is the average YSO lifetime \citep{evans09}, and $M_\emr{cloud}$ is the cloud mass, obtained by integrating the column density maps above a threshold of $1.105\e{21} \Htpscm$ (see Sect. \ref{sec:dense} for details).

The star formation properties for each cloud are summarised in Table \ref{tab:clouddata}. The YSO positions for all clouds are plotted in the supplementary online material (\href{https://zenodo.org/records/14360623/files/YSO_all.png}{YSO\_all}).

\clouddata{}


\section{Results} 
\label{sec:results}

\subsection{Volume densities in nearby molecular clouds}
\subsubsection{Volume density distributions} 
\label{sec:voldens}

We applied \coltt{} to the entire sample of 25 clouds. 
The reconstructed volume densities range from $1.3\Htpccm$ in the diffuse background of Orion\,A, to $1.3\e6\Htpccm$ in the densest core identified in Oph\,L1688. The lowest reconstructed volume densities depend both on the (low) column densities in a given field and on the total area of the field (and hence the reconstructed maximal depth along the line of sight); it is thus no surprise to find the lowest reconstructed density in the second largest map in our sample. We also note the presence of values higher than $1.3\e6\Htpccm$ in the Lupus clouds, but these are single-pixel noisy values at the edges of the maps, whereas the dense cores in Oph\,L1688 are robust structures, observed with a high \SNR{} and displaying a reliable morphology. All gas densities are expressed in units of H$_2$ to be consistent with the (dust-derived) HGBS datasets, but at the lowest column and volume densities one should note that not all hydrogen is expected to be present in the form of  H$_2$ -- however the H/H$_2$ fractionation is beyond the scope of this paper.

For the purpose of this pilot study, we only used the compressed version of the results, that is, the peak volume density maps and the global volume density PDFs. An example of these results is presented and compared with the original column density data in Fig. \ref{fig:panels} for the case of the Cep1228 cloud -- the rest of the cloud sample is presented in the \href{https://zenodo.org/records/14360623}{supplementary online material}.

\FigPanels{}

Common features emerge when comparing the volume density with column density in these figures. The maps of peak volume density (maximum volume density along the line of sight) show the concentrations of gas fragmented into a number of cores, and these cores are immersed in a common low-density envelope. On top of this pattern, and at almost all densities (but most visible at the lowest densities), one can see the presence of small, high density clumps. While it is tempting to attribute these clumps to noisy fluctuations of the column density maps, their aspect is different from the effect of added noise observed when test \coltt{} on simulations (Fig. \ref{fig:app:noiseone} and \ref{fig:app:noisetwo}), it is therefore likely that they reflect the real, clumpy structure of the ISM at all densities. The features of the column density PDFs (secondary peaks, slope breaks...) are for the most part reproduced in the volume density PDFs, but often appear smoothed out. A general exception can be noted regarding features at column densities lower than the main peak of the PDF, which are not found in volume density -- this is most likely a sampling effect. The tail of the volume density PDFs is in general flatter and longer than the one of column density PDFs, which corresponds to a larger dynamic range in volume than in column densities (see Sect. \ref{sec:discusscorrel}).

\subsubsection{Correlation of peak volume density with column density} 
\label{sec:correlation}

The joint distribution of the column and peak volume density (Fig. \ref{fig:panels}, lower right) presents a characteristic shape for all the clouds, which can be described as a ``feather''. A very tight correlation is found in a main ``stem'', which acts as a lower limit for the peak volume density for any column density, and extends throughout the entire data range. Above a certain volume density, ``barbs'' start to branch off from the stem; each of these barbs also traces a very tight correlation, some of which can branch again later. In the end, the cloud of points which represents the joint histogram of peak volume density vs. column density for the entire cloud is actually the superposition of a large number of line-like correlations of various statistical weights which get superimposed on top of each other.

This structure arises from the hierarchical reconstruction of the volume density. Each line-like correlation corresponds to a single region containing nested column density contours. As long as the successive contours are connected, they yield a single value of area, thus depth, and therefore a one-to-one correspondence between column density and volume density. But as soon as a contour break into two or more sub-regions, the one-to-one correspondence is lost and the correlation branches off. In this context, the main stem corresponds to the largest sub-region for each column density contour, it thus ranges from the contour enclosing the entire field to the contour containing the highest column density value in the map. This behaviour for example is very clear in the case of a small field with a very simple morphology such as the B68 globule (supplementary online material, \href{https://zenodo.org/records/14360623/files/4panel-B68.png}{B68}).\\

The relation between column density and peak volume density is particularly interesting as it allows in principle to use the column density to infer the highest volume densities reached along any line of sight - these highest densities being the ones involved in star formation. We therefore tried to characterise the average relation between column and peak volume density for the entire cloud sample. The joint distribution of column densities and peak volume densities for all the fields combined is shown in Fig. \ref{fig:colvol}.

\FigColVol{}

The main stems of the joint distributions of the individual clouds come out clearly as bright lines almost parallel to each other at low densities. A group high volume density points draws the eye in the shape of the distribution, but these are negligible in terms of statistical weight and correspond to noisy pixels at the edge of some of the column density maps. What is more significant is the lower edge of the distribution, which shows a change of slope around $5-10\e{21}\Htpscm$. Given that the majority of the data points are close to this lower edge, we have tested if the overall correlation reflects this change of slope. We therefore fitted this correlation with a power-law, over the entire data range, and in two limited column density ranges, corresponding to the commonly used definition of low-density and high-density molecular gas (see Sect. \ref{sec:dense}, namely $\Av = 1-8\magn$ and $\Av > 8\magn$ -- in our case converted to $N = 1.11\e{21} - 8.84\e{21}\Htpscm$ and $N > 8.84\e{21}\Htpscm$ using the conversion factor $N_\emr{H}/\Av = 2.21\e{21} [\pscm/\magn]$ \citep{guver09}; the threshold set at $\Av = 8\magn$ corresponds visually to the location of the change of slope. The results of the fitting are visualised in Fig. \ref{fig:colvol} and detailed in Table \ref{tab:colvol}. The very small errors in the fit results do not indicate a tight correlation -- the measured root mean square (RMS) scatter in indeed large -- but merely that the fit is well-constrained by the very large number of data points.

\TabColVol{}

The fit of the entire density range is dominated by the low to intermediate density material, which vastly dominate the cumulative area of the studied sample of clouds. On the other hand, the change of slope is well visible in the power-law exponents of the two-part fit, with a low-density exponent of $\sim 0.9$ and a high-density exponent of $\sim 2.2$. This suggests a difference in physical conditions, or at least in cloud morphology, between the two density regimes, and is further discussed in Sect. \ref{sec:discusscorrel}

\subsection{Star formation versus column and volume density} 
\subsubsection{Measuring the dense gas} 
\label{sec:dense}
Star formation requires dense gas to proceed, but it is necessary to quantify what this dense gas means, and whether the density requirement is absolute \citep[gas above a certain density threshold can contribute to star formation, \eg][]{gao04, lada10}, or relative \citep[only a top fraction of the gas in clouds can contribute to star formation, independent from the mean density of the parent cloud, \eg][see \citealt{padoan14} for a review]{kruijssen14, spilker21}. We therefore use two different metrics to empirically describe the density distribution in the studied clouds -- the dense gas fraction and the density contrast -- and apply them to both the column density PDF and the volume density PDF of each cloud.

The dense gas fraction, initially introduced by \citet{lada10}, is a commonly used descriptor of the amount of gas effectively involved in the star formation process via gravitational collapse. The dense gas fraction is defined as the ratio of the mass of the dense gas to the mass of the bulk of the cloud. The density contrast, introduced by \citet{spilker21}, provides instead a relative measure of the gas density, based on the properties of the individual column density PDFs rather than on predefined thresholds. This density contrast is defined between a low density density corresponding to the peak of the density PDF, and a high density which is the one above which 5 percent of the mass of the cloud resides (this mass being considered only above the low-density limit). Both these descriptors also present the advantage of not relying on a model of the PDF shape (\eg{} a power-law or a log-normal distribution), which would not necessarily fit the data well \citep{spilker21}

In practice we adopt the following definition for the column-density based  and volume-density based dense gas fraction ($f_\emr{dg}^N$ and $f_\emr{dg}^\rho$) and density contrast ($\Delta N$ and $\Delta \rho$):
\begin{equation}
    f_\emr{dg}^N = \frac{\int_{N_\emr{dense}}^\infty \!N\cdot P(N)\diff N}{\int_{N_\emr{cloud}}^\infty \!N\cdot P(N)\diff N}
\end{equation}
where for all clouds the thresholds are $N_\emr{cloud} = 1.11\e{21} - 8.84\e{21}\Htpscm$ and $N_\emr{dense} = 8.84\e{21}\Htpscm$, based on the commonly used extinction thresholds $\Av = 1$ and $\Av =8$ \citep{lada10,kainulainen13,evans14,spilker21}, and with a conversion factor of $N_\emr{H}/\Av = 2.21\e{21} [\pscm/\magn]$ \citep{guver09}; similarly, we have

\begin{equation}
    f_\emr{dg}^\rho = \frac{\int_{\rho_\emr{dense}}^\infty \! \rho \cdot P(\rho)\diff \rho}{\int_{\rho_\emr{cloud}}^\infty \!\rho\cdot P(\rho)\diff \rho}
\end{equation}
where for all clouds the thresholds are $\rho_\emr{cloud} = 2\e2\unit{H_2\pccm}$ and $\rho_\emr{dense} = 2\e4\unit{H_2\pccm}$, based on the results of tests described in more details in Sect. \ref{sec:thresholds} following and approach similar to what \citet{lada10} used to derive the column density thresholds;

\begin{equation}
    \Delta N = \log_{10} \frac{N_{5\%}}{N_\emr{peak}} \;\; \emr{and} \;\; \Delta \rho = \log_{10} \frac{\rho_{5\%}}{\rho_\emr{peak}}
\end{equation}
where $N_\emr{peak}$ and $\rho_\emr{peak}$ are respectively the column density and the volume density values corresponding to the peak (most probable value) of their respective PDFs, and the ``top 5\%'' values $N_\emr{5\%}$ and $\rho_\emr{5\%}$ are defined by:
\begin{equation}
    \frac{\int_{N_\emr{5\%}}^\infty \!N\cdot P(N)\diff N}{\int_{N_\emr{peak}}^\infty \!N\cdot P(N)\diff N} = 0.05  \;\; \emr{and} \;\;  \frac{\int_{\rho_\emr{5\%}}^\infty \!\rho\cdot P(\rho)\diff \rho}{\int_{\rho_\emr{peak}}^\infty \!\rho\cdot P(\rho)\diff \rho} = 0.05
\end{equation}

The determination of the value of $N_\emr{peak}$ and $\rho_\emr{peak}$ from a PDF which is in practice a discrete histogram is very sensitive to sampling, which can change the limits of bins and thus affect the number of points in each bin. To alleviate this effect, we used a constant number of bins for all the column and volume density histograms (100 bins with logarithmic sampling), and we also smoothed the obtained histograms with a 3-bin wide sliding average (Fig. \ref{fig:panels}, lower left). The peak value used as $N_\emr{peak}$ (or $\rho_\emr{peak}$ respectively) was then extracted from these smoothed histograms.

\TabDense{}

The values of $f_\emr{dg}^N$, $f_\emr{dg}^\rho$, $\Delta N$ and $\Delta \rho$ for all the clouds in our sample are presented in Table \ref{tab:dense}.\\

We compare for each cloud the star formation efficiency (Eq. \ref{eq:sfr-sfe}) derived from the YSO counts described in Sect. \ref{sec:yso}. The choice of the SFE rather than SFR allows for a better comparison between the clouds, given than the SFE is, like all the dense gas descriptors, normalised by cloud mass. The correlations between the SFE and $f_\emr{dg}^N$, $f_\emr{dg}^\rho$, $\Delta N$ or $\Delta \rho$ were then quantified by performing a fitting. The fitting was linear in each case, and was performed in log-log space in the case of SFE vs. $f_\emr{dg}^N$ and $f_\emr{dg}^\rho$, and in lin-log space in the case of SFE vs. $\Delta N$ and $\Delta \rho$, since the density contrast is already defined as a logarithmic quantity. Considering the star formation efficiency in log space also effectively excluded from the correlation the B68, Cep1241 and Polaris fields, the SFE of which is 0. The tightness of the correlation was in each case estimated both in terms of the fit quality ($R^2$, or error on the fit parameters), and it terms of RMS scatter with respect to the fitted linear relation. These correlations and the fitting results are presented in Fig. \ref{fig:sfe} and in Table \ref{tab:sfe}. The narrowest scatter is obtained for the column-density based density contrast (0.197\,dex), while the volume-density based dense gas fraction not only yields a RMS scatter almost as tight as the best one (0.197\,dex), but has also by far the best fit, making it overall the best available predictor of star formation.

\FigSFE{}

\TabSFE{}

\subsubsection{Volume density threshold for star formation} 
\label{sec:thresholds}

Translating the column-density based definition of the density contrast $\Delta N$ to volume densities is straightforward (owing to the data-based nature of this descriptor). Adapting the definition of the dense gas fraction $f_\emr{dg}$ for the volume density case, however, requires defining \emph{ad hoc} low- and high-density thresholds. Based on our sample and following an approach close to the one initiated by \citet{lada10}, we obtained values of $30\Htpccm$ and $2\e4\Htpccm$ low- and high-density thresholds respectively, which mirror the $\Av = 1$ and $\Av = 8$ extinction thresholds used in the column density case.

The high density threshold corresponds to the density above which the gas reservoir is the most directly involved in star formation. This was quantified by \citet[their Fig. 3]{lada10} by measuring the correlation between the number of YSOs in a sample of clouds, and the mass of gas above a given density threshold. The tightest correlation was found for a threshold of $\Ak = 0.8 \pm 0.2$, corresponding to $\Av = 7.3 \pm 1.8$. 

The low density threshold is used to separate the bulk of the molecular cloud from its diffuse fore- and background. This enables normalising the relation between star formation and gas reservoir by the cloud mass, which becomes a relations between SFE and dense gas fraction, and therefore makes it possible to compare clouds of different sizes. The commonly used value of $\Av = 1$ stems from the $\Ak = 0.1$ threshold introduced for that purpose by \citet{lada12}, it is also close to the extinction levels at which molecular ($^{12}$CO) emission starts to be detected \citep[\eg][]{goldsmith08,pety17,lewis21}.

For volume density thresholds, we assume in practice that the low density threshold should be approximately the volume density at which the gas becomes molecular, so of the order of $\sim 100 \unit{\emr{H}_2\pccm{}}$ \citep{heyer15, pety17}. The high density threshold, on the other hand, could in principle be anywhere between the density at which the gas typically get structured into a filamentary network \citep[$\sim 1\e3\Htpccm$,][]{pineda10,orkisz19}, and the densities reached in prestellar cores \citep[$5\e4 - 1\e6 \Htpccm$, \eg][]{evans09,teixeira16}.
We therefore test a range a value for both the low density threshold ($1\e1 - 5\e2\Htpccm{}$) and the high density threshold ($1\e3 - 1\e6 \Htpccm{}$). The resulting scatter in the correlation between $f_\emr{dg}^\rho$ and SFE is presented in Fig. \ref{fig:scatter}.

\FigScatter{}

One can observe that varying the low density threshold does a minor effect on the tightness of the correlation, as shown by the fluctuation in position of the transparent blue plots in Fig. \ref{fig:scatter}. The main variation comes from the value of the high density threshold, and the minimum scatter is obtained at (or very close to) the same value of high-density threshold, namely $2\e4 \unit{\emr{H}_2\pccm{}}$, for all values of low-density threshold. The high-density threshold of $2\e4 \unit{\emr{H}_2\pccm{}}$ is therefore a robust result in terms of producing the smallest scatter in the correlation between $f_\emr{dg}^\rho$ and SFE. The absolute minimum scatter is obtained with a low-density threshold set at $2\e2 \unit{\emr{H}_2\pccm{}}$. 

We also note that scatter values for high density thresholds beyond $5\e4 \unit{\emr{H}_2\pccm{}}$ (shaded area in Fig. \ref{fig:scatter}) are not meaningful, because several clouds in our sample do not reach such high volume density values, while still exhibiting star formation. Higher density thresholds for star formation not only produce poorer correlation measurements due to the diminishing number of clouds in the sample, but would also in contradiction with the presence of active star formation in the clouds not reaching these densities. 

\section{Discussion}
\label{sec:discussion}
\subsection{Overview of limitations} 
\label{sec:discusslimit}
The quality of our results is subject to two independent sets of limitations, one coming from the data used, and the other from our data processing, which means the \coltt{} algorithm and the way it is applied to the data.\\

Regarding the data used, we made the choice to favour homogeneity over quality.

As explained in Sect. \ref{sec:yso}, the YSO catalogues that we use are not the most recent and arguably most complete available. This has implications in terms of completeness and reliability of the YSO sample -- for example, the YSO catalogue for the Orion complex produced by \citet{megeath16} using additional \emph{Chandra} X-ray data contains 408 more YSOs than the \citep{megeath12} catalogue that we are using, and including completeness corrections it is estimated that the Orion complex contains a total of 5104 YSOs, rather than the 3481 we used as a starting point. However, while incomplete YSO catalogue do affect the measurements of star formation rates, the overall scaling between star formation and dense gas should not be affected significantly: indeed, the homogeneity of the YSO catalogues ensures that the completeness fractions should be approximately the same for all clouds, and therefore the slopes of the correlations we measured should be unchanged.

The column density data used pose another question. While extremely homogeneous in terms of observations and data reduction, given that all the maps were obtained in the same observational survey, the column density maps are arguably not the best possible that could have been obtained using the \emph{Herschel} data. Other, more refined ways of computing total column densities from SEDs than pixel-by-pixel modified black-body fitting do exist, like for example the PPMAP \citep{marsh16} method. Besides, the column density values obtained by the HGBS consortium from their observational data do not necessarily agree with other estimations of column density for the same clouds, obtained either from other observations (notably extinction data) or even using the same \emph{Herschel} data, as can for example be seen in \citet{konyves20} and \citet{lombardi14}. Re-deriving column densities is beyond the scope of this study, hence our use of the published data as they are. While the homogeneity of the column density data-set ensures that the measured correlations with volume densities and with star formation properties remain valid, absolute values such as the volume density threshold for dense gas could be shifted if a different derivation of column densities had been used.\\

The selection of fields to be passed to \coltt{} also raises questions, on two particular points: noise, and completeness.

The question of noise was also faced by the HGBS team when studying these clouds, and is mostly restricted to the edges of certain maps (Cepheus Flare and Lupus in particular). The solution adopted by \eg{} \citet{difrancesco20} was simply to manually draw a border separating the noisy edge of the map from the rest of the field. For simplicity's sake, we avoided any manual definition of the fields to be retained, and only applied the filtering and erosion  described in Sect. \ref{sec:coldens}, however some noisy pixel with extreme column density values were left. Because they are outliers, their presence is very visible in plots such as Fig. \ref{fig:colvol}, but given the very limited number of these pixels their contribution to the mass statistics, to the correlation measurements and to the volume density estimation is negligible.

The matter of completeness arises mostly through the question of close contours of column density \citep{alves17}, that is whether any given column density contour is fully included in the observed field or if it is cropped at the edges. In our context it translates into an uncertainty, and even a systematic error, in the estimation of the volume associated with the lowest column density contours in the studied clouds. However, due to the combination of the facts that the incomplete contours correspond to the lowest column densities and to the largest areas (almost the entire observed field), the resulting volume densities are very low, and the error which affects them is therefore negligible. Say otherwise, an error on volume estimation even by a factor of a few corresponding to background volume densities of a few tens of \Htpccm{} do not affect in any significant way the statistics derived from the volume densities of dense regions at $\rho \gtrsim 1\e4 \Htpccm$.\\

The main limitations in the volume density estimations come however from the assumptions implemented into the \coltt{} code, and the inherent errors in the reconstruction which are inevitable in the lack of a real knowledge of the three-dimensional structure. The assumptions spelled out in Sect. \ref{sec:geometry} lead, as any model, to some amount of oversimplification. The isotropy hypothesis is, in the lack of any better knowledge, a safe choice. It can be a wrong assumption for individual clouds, as illustrated by \citet{rezaei22}, but at the scale of a sample of clouds, the statistical properties are largely independent from the cloud orientation \citep{kainulainen22}. As an additional complexity, while on scales of tens of parsecs some clouds are known to be anisotropic, this aspect ratio cannot be expected to remain the same at all scales: in the densest regions one would expect to find approximately cylindrical filaments with diameters of the order of 0.1\pc{} and lengths of up to several parsecs \citep[\eg][]{orkisz19,suri19}, as well as roughly spherical prestellar cores of about 0.1\pc{} diameters \citep[\eg][]{kirk16,arzoumanian19} -- independently of the large-scale aspect ratio of the parent cloud. Besides, in \coltt{} isotropy is currently implemented in a rigid and uniform way, by ascribing the same depth to all points within a contour. A development of the volume density estimation could thus be to take into account the shape and local aspect ratio of the contours, while keeping a global, statistical isotropy of the volume density distribution. Steps in that direction have for example been taken by \citet{hasenberger20}.

The hypothesis of hierarchically nested contours (and volumes) is in that sense a continuation of our approach to isotropy: if a contour is fragmented into independent regions in the plane of the sky, its 3D  reconstruction is also likely to be fragmented along all axes. But, faced with the impossibility of determining where and how this fragmentation along the line of sight might happen, it is omitted in favour of a simpler, hierarchical solution.

In that view, \coltt{} shows that even a simple reconstruction of volume densities can be a valuable tool for the study of the ISM, but further research into better ways of inferring volume densities from plane-of-the-sky observations is of course necessary.

\subsection{Column densities vs. volume densities}
\label{sec:discusscorrel}

In comparing the column densities with the obtained volume densities, a first aspect to consider is the relation between the PDFs of the two quantities for each molecular cloud. \citet{brunt10a,brunt10b} propose a statistical model for reconstructing directly the volume density PDF based on the observed column density PDF, where the $\rho$-PDF is simply a shifted and scaled version of the $N$-PDF, and the scaling in width depends on the power spectrum of the column density PDF. We can see that this model is close to what we see on the lower-left panel of Fig. \ref{fig:panels} (as well as in the \href{https://zenodo.org/records/14360623}{supplementary online material} for all other clouds). As mentioned in Sect. \ref{sec:voldens}, the volume density PDFs are indeed very similar in shape to the column density ones. Regarding the width scaling, the ratio of the $\rho$-PDF width to the N-PDF width is $1.78 \pm 0.23$ dex  -- while \citet{brunt10a} find a ratio between 2 and 3 when studying simulations, 2 corresponding to purely compressive turbulence. The observed spread is narrow, but one can however notice in Fig. \ref{fig:width} a physical trend in the relation between the $N$-PDF and $\rho$-PDF: with the exception of the outliers Cep1228 and Serpens, the $N$-PDFs of which have unusually extended tails at low column densities, the ratio of the PDF widths seems to anti-correlate with the SFE of the clouds. The expected trend of increased star formation correlating with wider PDF, characterised by a broader, compressive log-normal part and a heavy self-gravitating tail \citep[\eg][]{klessen00, federrath13, burkhart19} is therefore less present in the $\rho$-PDFs than in the $N$-PDFs -- whether this is a physical effect or a methodological bias will require future investigations.

\FigWidth{}

The other aspect to examine is the relation illustrated in Fig. \ref{fig:colvol} between the column density and the peak volume density along each line of sight. This relation, when considered at the scale of the entire cloud sample, yields two different regimes, one at low densities where $\rho_\mathrm{peak} \sim N$ and one at high densities where $\rho_\mathrm{peak} \sim N^2$, the transition between the two occurring around the often used ``dense gas threshold'' of $A_\mathrm{V} = 8\unit{mag}$. The transition between the two is however quite smooth, and can be understood in the light of the ``feather'' shape of the individual joint distributions of column density and peak volume density in the lower-right panel of Fig.  \ref{fig:panels} (as well as in the figures of the \href{https://zenodo.org/records/14360623}{supplementary online material} for all other clouds). As explained in Sect. \ref{sec:voldens}, the ``feather'' traces the fragmentation of the cloud into an increasing number of substructures. As long as the contours of increasing column density do not fragment, their area diminishes slowly to the point where it can be considered as almost constant, hence the proportionality between $\rho_\mathrm{peak}$ and $N$. When fragmentation occurs, however, the volume corresponding to the total area $A_i$ becomes $\sim A_i^{3/2}/f$, where $f$ is the number of fragments -- reconstructed volume densities therefore increase abruptly by a factor $\sim f$. The joint distribution splits into a number of ``barbs'' and their slope increases sharply. One can even see in some cases that, if no further fragmentation occurs at higher densities, the barbs can flatten out again -- a persistently increased slope of the relation between $\rho_\mathrm{peak}$ and $N$ requires that fragmentation continues at increasing densities. The column densities at which major fragmentation occurs vary from cloud to cloud, but the global relation seems to indicate that in general fragmentation become more prevalent in molecular clouds above column densities of $\sim 5 - 10 \e{21} \Htpscm{}$.

\subsection{Comparison with other works}
\label{sec:discusssfr}

\subsubsection{Impact of volume density on star formation rates}

There is an exhaustive number of works studying the possible column density threshold for star formation in cores \citep[\eg][]{elmegreen02} or clouds \citep[\eg][]{lada10, evans14}, and the column density distribution is still largely considered as a reliable predictor of star formation \cite[]{retter21}. Similarly, many works relate the column densities to volume densities through a simplistic assumption for geometry, usually a sphere, which then enables considering the threshold in terms of volume density. For example, \citet{lada10} derives an extinction threshold of $A_K = 0.8 \magn$ and from that estimates a volume density threshold of $10^4 \Htpccm$, very similar to what we obtain in the present work.

Another approach to the study of the relation between the volume density of molecular gas and star formation is based on the observation of so-called dense gas tracers, such as the 3-mm band transitions of HCO$^+$, HNC or HCN. In the wake of the work of \citet{gao04} on HCN, these molecular lines have been used, due to their high critical densities ($\rho \sim 1\e4 - 1\e5 \Htpccm$, to measure the amount of gas present above these densities and compare it with various tracers of star formation, in molecular clouds or nearby galaxies \citep[\eg][]{bigiel16,jimenez19,kauffmann17}. However, it is likely that the close correspondence between the volume density threshold for star formation found in the present study and the densities allegedly traced by HCN$(J=1-0)$ is a mere coincidence. Indeed, the critical density is not a hard threshold for the emission of molecular lines, so that not only there is a large spread in the estimation of the volume densities sampled by these tracers (from $\sim 1\e4$ to almost $1\e6 \Htpccm$), but there is an increasing amount of arguments questioning whether ``dense gas tracers'' do actually trace specifically the dense gas content of molecular clouds \citep{pety17,shimajiri17,tafalla21,dame23}, given that HCN is actually detected over a very broad range of column (and thus volume) densities in molecular clouds -- this is among others due to the fact that HCN is mostly observed to be optically thick while its critical density decreases with opacity \citep[\eg][]{shirley15}. In particular \citet{tafalla21} argue that HCN$(J=1-0)$ rather traces the total gas content of clouds, its very linear correlation with column density being a lucky interplay between excitation and chemical abundance effects. In that context, it should be understood that the density threshold for star formation of $2\times 1\e4\Htpccm$ derived from the \coltt{} volume density reconstructions is a fully independent result, which provides no argument in favour of the use of HCN, HNC or HCO$^+$ lines to trace the star-forming gas in molecular clouds. Conversely, the value of this threshold does not invalidate the use of specific astrochemical tracers such as N$_2$H$^+$ to trace even denser gas in clouds and cores \citep[\eg][]{kirk16,pety17,kauffmann17}.

An interesting approach, which takes a step towards volume density modelling, has been pioneered by \citet{hu21}. The authors acknowledge that column density measurements are not sufficient to determine accurately the star formation activity, and in particular the free-fall time of molecular clouds. However, rather than trying to directly reconstruct the volume density distribution, they use simulations to build a model which indirectly derives the cloud free-fall time (and therefore the star formation rate) from the column density distribution. A direct comparison with our study based on volume densities is thus unfortunately not possible.

Even more relevant to the present case, \citet{kainulainen14} presented a method to derive volume density PDFs of molecular clouds via inverse modelling and applied it to a sample of nearby clouds.

Based on their volume density data, \citet{kainulainen14} derive a star formation threshold of $5\e{3} \pccm$. It is important to recall that \citet{kainulainen14} defined star formation threshold as the highest densities in clouds that had no or very little star formation. The threshold we derive in this paper is significantly higher, at $2\e{4}\Htpccm$. However, this difference is likely related not only to differences in definitions of the star formation threshold, but also differences in data sets (spatial resolution, dynamic range, and star formation numbers). In addition to the more limited column densities probed by their extinction data, \citet{kainulainen14} map their cloud at a coarser resolution of 0.1\pc{}. For a volume density reconstruction with \coltt{}, a lower resolution has the double effect of reducing the column density peaks by washing them out, and of making the size of the inferred volumes larger -- both effects contribute to lowering the reconstructed volume densities, with the former being largely dominant. Figure \ref{fig:resolution} shows the maximum volume densities and SFE in the clouds that overlap between our and \citet{kainulainen14} samples, comparing the cases of the $36.3\arcsec$ \emph{Herschel} resolution and of a smoothed, $0.1\pc{}$ resolution. The maximum volume densities are heavily affected by resolution, and in particular, many clouds gather around $1-2 \times 10^{3}$ cm$^{-3}$ in the smoothed resolution. This order-of-magnitude effect is the likely reason for the difference between the star formation thresholds we and \citet{kainulainen14} derive. Using the \coltt{} volume density reconstruction along with the \citet{kainulainen14} threshold definition would yield a density threshold of $1 - 1.5\e{3}\Htpccm$ and $6 - 10\e{4}\Htpccm$ at $0.1\pc{}$ and $36.3\arcsec$ resolutions respectively, showing that these different approaches yield compatible results.

\FigResolution{}

\subsubsection{Volume density inference}

The work of \citet{kainulainen14} described above is particularly interesting in its attempt to reconstruct volume densities, despite the limitations of the column density data used, and the lack of focus on the spatial distribution (maps) of volume densities.

In that context, another inversion approach for determining volume densities for a generic cloud geometry has been presented by \citet[][]{hasenberger20}. Their method, named AVIATOR, is based on inverse Abel transform of column density data. \citet{hasenberger20} did not systematically apply their method to full molecular clouds, but tested it in the case of two relatively simple globules which are part of our sample, B68 and L1689B, and compared the results with the dedicated modelling results obtained for these globules by \citet{roy14}. Since the column densities derived in these works differ significantly from the ones derived by the HGBS consortium and used here, a direct comparison of the reconstructed volume densities in not possible. However, the ratios of volume to column density agree remarkably well for B68: the difference for the centre of the globule is of $+5\%$ and $-4\%$ with respect to \citet{roy14} and \citet{hasenberger20} respectively. In the case of L1689B, the match is still reasonably good, at $-6\%$ and $-23\%$ respectively -- this larger difference might be explained by the fact that our reconstruction was influenced by the complex structure of the extended environment of L1689B, while \citet{roy14} and \citet{hasenberger20} focused on the dense core only. It is however unclear how AVIATOR would perform in the case of more complex structures such as the ones tackled in the present study; applying it systematically would be an interesting avenue to studying the volume density structure of clouds.

The B68 and L1689B globules are ideal cases for volume density reconstruction, given their simple geometries, isolation which removes any complex contribution from the fore- and background, and lack of significant substructure at higher angular resolutions. Comparisons with volume density estimates in other, more complex environments can show the limitations of the \coltt{} method and of our data set. For example, densities measured at $\sim$0.1 pc scales in the filamentary structures of nearby, actively-star-forming molecular clouds range around roughly $10^{5} - 10^{6} \Htpccm$\citep[e.g., ][and references therein]{hacar23}. The maximum volume densities we infer for clouds like Orion\,A, CrA, and Taurus L1495 are in this ballpark. In case of less active filamentary clouds like Musca, dedicated works infer slightly lower densities of roughly $5-10 \e{4} \Htpccm{}$ \citep[][]{kainulainen16, cox16}; our estimates are in a reasonable agreement with that trend. Overall, it seems that at least in simple filamentary morphologies the estimates are in line with other studies, despite the fact that our model does not take the filamentary aspect into account.

As the geometry becomes more complex, it is expected that our inferred densities become more inaccurate, due to both fragmentation along the line of sight and anisotropy. For example, in the complex star-forming environments of the Aquila Rift and Orion A, dense cores with densities clearly higher than $10^6$ cm$^{-3}$ have been identified \citep[\eg][]{konyves15,shimajiri15}. Our maximum densities for Aquila and Orion A are about 0.2 and 0.6 $\times 10^6$ cm$^{-3}$, respectively. While still within the right order of magnitude, our estimates in this kind of environments become clearly more inaccurate due to the oversimplifying geometrical assumptions of the method.

Another limitation, discussed in the context of Fig. \ref{fig:resolution}, is the question of spatial resolution. Due to the modest angular resolution of \emph{Herschel} data, we cannot reach spatial scales (and therefore volumes) small enough to be able to probe the densest parts of prestellar cores (\eg{} the $\sim10^{7} \Htpccm$ reached in Orion\,A \citep{sahu23} derived from ALMA observations with an angular resolution up to 100 times better than \emph{Herschel}).

\section{Conclusion}
\label{sec:conclusion}

In order to study the role of the volume density distribution of molecular clouds in their star formation activity, we have developed and presented \coltt{}, an algorithm which performs a simple geometrical inference of the statistical distribution of gas volume density in a cloud based on the morphology of its column density map.

We applied \coltt{} to \emph{Herschel}-derived maps of column density for 25 nearby molecular clouds, and compared the properties of the obtained volume density distributions with the original column density data, and with the star formation properties of the individual clouds. We have observed the following significant trends:

\begin{itemize}
    \item The correlation between the column density and the peak volume density shows a piece-wise power-law behaviour, with two distinct regimes. At low column densities, the exponent of the power-law is $\sim 1$, at high column densities, this exponent becomes $\sim 2$. The transition between the regimes happens at column densities of the order of 5--10\e{22}\,H$_2$\pscm{}, which is compatible with the commonly adopted ``dense gas'' limit of 8 magnitude of \Av. The difference between these two correlation regimes is most likely an illustration of the importance of hierarchical fragmentation in dense gas.
    \item We tested two different dense gas descriptors (dense gas fraction and density contrast), and applied them to the column density and volume density distributions of the clouds. Out these four different measurements, the volume-density based dense gas fraction has the tightest correlation with the star formation efficiency of the clouds. This illustrates that star formation laws are controlled by physical quantities (such as volume density), not by observed quantities (such as column density), and underlines the importance of retrieving these physical quantities from observational data, even if indirectly.
    \item The correlation between the star formation efficiency and the volume-density based dense gas fraction (or between the star formation rate and the mass of dense gas) is tightest with a low density threshold (which separates the bulk of the cloud from the diffuse background) set at 2\e2\,H$_2\pccm$ and a high density threshold (which separates the dense, star-forming gas from the bulk of the cloud) set at 2\e4\,H$_2\pccm$. 
\end{itemize}

Many refinements in the treatment of the cloud morphology or in the aggregation of additional observational data could be implemented in order to improve the volume density reconstruction, but we have already shown with our results obtained with a purposefully simplistic approach, that reconstructing volume density has the potential of being an extremely powerful tool to study the ISM. Follow-up studies involving the volume density statistics, star formation properties and the chemistry of clouds using \coltt{} will further reveal this potential.

\section*{Data availability}
Supplementary figures, in particular the equivalents of Fig. \ref{fig:panels} for all clouds in the studied sample, are available at the following address: \url{https://zenodo.org/records/14360623}.

The \coltt{} code is referenced at the following address: \url{https://ascl.net/xxxx.xxx}, and available for downloading with its dependencies at the following address: \url{https://github.com/jan-orkisz/colume}.

\begin{acknowledgements}
This research has made use of data from the Herschel Gould Belt survey (HGBS) project (http://gouldbelt-herschel.cea.fr). The HGBS is a Herschel Key Programme jointly carried out by SPIRE Specialist Astronomy Group 3 (SAG 3), scientists of several institutes in the PACS Consortium (CEA Saclay, INAF-IFSI Rome and INAF-Arcetri, KU Leuven, MPIA Heidelberg), and scientists of the Herschel Science Center (HSC). Jan Orkisz acknowledges funding from the Swedish Research Council, grant No. 2017-03864. The authors thank Maryvonne Gerin for helpful comments on the writing of this paper.
\end{acknowledgements}

\bibliographystyle{aa} %
\bibliography{main} %

\newpage
\begin{appendix}
\label{app}
\section{Tests on simulated data} 
\label{app:simulations}
\subsection{Data and test procedure}
\label{ap:simdatatest}
Given that our volume density estimation algorithm is intrinsically assuming that the density distribution of clouds is hierarchically fragmented into centrally concentrated sub-regions, we have to choose astrophysical simulations which match this description reasonably well in order to obtain meaningful results. For that purpose, we selected the simulation data presented and studied by \citet{ibanez2017} and \citet{chira2019}, available as part of the CATS \citep{burkhart2020} database. This simulation is a zoom-in focused on star-forming clouds in a galactic context. A portion of galactic disc was simulated including ideal compressible magneto-hydrodynamics (MHD), gas heating and cooling, supernova driving and a static gravitational potential. Three dense clouds were selected in this environment and their collapse under the effect of self-gravity was followed at a higher resolution, with snapshots being taken every 0.1\Myr. The simulation was run using the FLASH \citep{fryxell2000,dubey2008} adaptive mesh refinement code.

The data used for our tests is the snapshot of a cloud with an initial mass $M = 4\e3 \Msun$, after 3\Myr\ of evolution under the effect of self-gravity. From this snapshot we extract a $40\times40\times40\pc$ cube of volume density values (expressed in H\pccm) with a adaptive resolution ranging from 0.4 to 0.1\pc. The volume density distribution (Fig. \ref{fig:app:pdf}) of the cloud is tri-modal, with most of the volume of the simulation cube corresponding to densities around 1\e{-3} and 1\e{-1}\,H\pccm{}.
Most of the mass of the cloud is however found in the last peak of the PDF, at densities higher than 1\,H\pccm{}. The densities are nevertheless not particularly high, not even reaching 1\e5\,H\pccm{}. This is compatible with the volume densities estimated for low-density filamentary networks \citep[\eg][]{orkisz19} or low-mass cores \citep[\eg][]{kirk16}, but filaments or dense cores can also reach densities an order of magnitude larger than this \citep[\eg][]{teixeira16}. The density distribution of this simulated cloud is therefore not perfectly representative of the most massive star-forming clouds in our sample.

In order to generate simulated data-sets, the simulated cube is simply summed along each of its axes to produce three column density maps. At this stage, in order to simulate the presence of a diffuse fore- and background, a uniform offset of 0.1\,H\pccm\ is added to the cube - this has the consequence of virtually removing the low-density peaks in the volume density PDF, and adds an offset of 1.23\e{19}\,H\pscm\ to the column density maps. The median column density value is 2.2\e{19}\,H\pscm\ and the maximum 3.6 -- 5.9\e{22}\,H\pscm\ depending on the projection.

The twelve (three clean, nine noisy) maps were then passed to the \coltt{} code with an arbitrary distance and the matching angular resolution, with 800 sampling levels with the standard inverse log percentile binning (Sect. \ref{app:sampling}). The maps of mean and peak density along the line of sight were generated, as well as volume density PDFs, and compared with the reference values of the noiseless simulated cube.

\FigAppPdf{}

\subsection{Results}
\label{app:simresults}
\subsubsection{Reliability of the density estimation in the ideal case}
\label{app:simreliability}

The noiseless column density maps produced from a simulated cloud isolated within a cube are the ideal scenario in which to run a volume density reconstruction. Real observational data are be noisy, the fore- and background contamination as well as the limited observational field make it difficult if not impossible to define clearly the boundaries of the cloud, not to mention our ignorance in most cases regarding the isotropy of the cloud. However, this ideal case is still a necessary test-bed for the \texttt{Colume} code.

By construction, the \texttt{Colume} method is conservative in terms of total mass, down to sampling precision -- this is confirmed in this test with a precision of $\lesssim 10^{-5}$. In the particular case of this test, this is also true on a pixel-by-pixel basis: since the shape of the original object is really a cube, the geometry of the outer envelope of the cloud is perfectly reconstructed.

Figure \ref{fig:app:cleanp} shows the quality of reconstruction of the peak volume density reached along each line of sight for the $x$-axis projection of the cloud (the $y$ and $z$ projections show similar behaviours). We can see than the reconstructed peak density is typically underestimated by up to an order of magnitude, however this bias is not uniform throughout the density range. As it can be seen from the upper right and lower left panels of Fig. \ref{fig:app:cleanp}, the reconstructed peak density is very close to the original for very low ($\lesssim$ 1\,H\pccm) and high ($\gtrsim 10^4$\,H\pccm) volume densities. At intermediate densities, it underestimates the peak volume densities much more. This can be understood in terms of scale and structure of the gas (which is well illustrated by the lower left panel of Fig. \ref{fig:app:cleanp}). The lowest and highest volume densities correspond to the envelope of the cloud and to dense cores respectively, which have rather isotropic shapes. On the other hand, at intermediate densities, the gas is mostly distributed into sheets and filaments, for which the isotropy assumption which underlies our volume reconstruction in inappropriate -- the estimated volumes are too large, and consequently the densities too low. This scale-dependent effect is present for all the density increments encountered along a given line of sight, and the peak volume density, being the sum of all these increments, displays this bias in a particularly visible way.

\FigAppCleanP{}

The volume density PDF is also a crucial element in order to derive quantity such as the dense gas fraction or density contrast. Figure \ref{fig:app:pdfxyz} shows that the reconstruction of this PDF is very consistent no matter what the projection angle is. 

\FigAppPDFxyz{}

\subsubsection{Influence of noise}
\label{app:simnoise}

In order to test the behaviour of the \texttt{Colume} code when dealing with imperfect data, we added three different patterns of Gaussian noise to the clean column density maps. The first is a noise with a uniform noise amplitude of 3.7\e{19}\,H\pscm{}, corresponding to three times the background level - the \SNR therefore varies between 1/3 and $\sim 1000$ throughout the maps. The second pattern is a constant signal-to-noise of 10 (Gaussian amplitude proportional to the signal). The third pattern is the sum of the previous two.

Figures \ref{fig:app:noiseone} and \ref{fig:app:noisetwo} show the comparison between the reconstructed volume densities and the original (noiseless) data.

\FigAppNoiseOne{}
\FigAppNoiseTwo{}

The uniform noise (Fig. \ref{fig:app:noiseone}) heavily affects the regions of low density, but this has consequences also on high density regions despite their very high \SNR. What happens is that regions which otherwise would be smooth are disrupted by noise on a pixel-by-pixel basis, each noisy pixel being effectively treated as a small small dense core. This results in many low density pixels being incorrectly reconstructed as (very) high mean and peak densities, which is obvious in the joint histograms of Fig. \ref{fig:app:noiseone} (panels 2 and 5). 

\afterpage{\clearpage}

\section{Column density sampling} 
\label{app:sampling}
As described in Sect. \ref{sec:implementation}, the size of the generated volume density data-set is of the order of $X\times Y \times S$, where $X \times Y$ are dimensions of the column density map and $S$ is the number of column density increments. The native number of increments, $S_0$, is equal to the number of unique values in the column density map, with $S_0 \leq X \times Y$ -- in practice with high-precision data and in the presence of noise $S_0 \sim X \times Y$. The native maximum size of the volume density data-set is thus of the order of $\lesssim \left(X \times Y \right)^2$. Such a size in unmanageable in terms of random-access memory requirements during the computation: once accounted for the different data by-products, buffers, and for the digital precision used for the data, even a moderate field such as the $400\times400$ pixel test filed used in Appendix \ref{app:simulations} requires several terabytes of RAM. A sampling such that $S \ll S_0$ is therefore necessary, with the constraint of recovering as much as possible of the detailed cloud structure.

Two main approaches are taken to sample the column density values: a ``min-max'' approach, which is agnostic to the distribution of column density values except for its minimum and maximum, and a ``percentile'' approach, which as the name suggests follows the percentiles of the distribution (note that we consider the distribution of unique column density values, rather that the column density PDF of the entire field). In both cases, the spacing of the sampling bins can be set in different ways, the ones we implemented are the following: a linear spacing (where the difference between successive bins is constant), a conventional logarithmic (log) spacing (where the ratio between successive bins is constant), and an inverse logarithmic (i-log) spacing, which is the base-10 logarithm of a linear spacing, rescaled to the range of the data. For a number of samples $S$, with extrema $N_\mathrm{min}$ and $N_\mathrm{max}$, and percentiles of distribution $Y$ being noted $p_x(Y)$ (with $x$ between 0 and 100), the $i$-th bin threshold $N_i$ is thus defined by:

\begin{itemize}
    \item Min-Max linear
    $$ N_i = N_\mathrm{min} + \frac{i}{S} \left(N_\mathrm{max}-N_\mathrm{min}\right) $$
    \item Min-Max log
    $$ N_i = N_\mathrm{min} \cdot \left(\frac{N_\mathrm{max}}{N_\mathrm{min}}\right)^{i/S} $$
    \item Min-Max i-log
    $$ N_i = N_\mathrm{min} + \left(N_\mathrm{max}-N_\mathrm{min}\right) \cdot \log_{10}\left(1+ i/S \cdot 99 \right)/2 $$
    \item Percentiles linear
    $$ N_i = p_{x_i}(N),\; x_i = i/S\cdot100 $$
    \item Percentiles log
    $$ N_i = p_{x_i}(N),\; x_i =\left(\frac{100}{100/S}\right)^{i/S}\cdot\left(\frac{100}{S}\right) $$
    \item Percentiles i-log
    $$ N_i = p_{x_i}(N),\; x_i = 100\cdot\log_{10}\left(1+ i/S \cdot 99 \right)/2 $$
\end{itemize}

The definitions of the ``i-log'' scales and of the ``percentiles log'' scale are somewhat arbitrary, due to the impossibility to define in a unique,  natural way a logarithmic function which maps the $[0,1]$ range on itself. The ``i-log'' definition is thus chosen to be the most natural for percentiles, and the ``percentiles log'' scale is define so that its first bin above 0 has the same value as in the ``percentiles linear'' sampling.

The bins created by six different sampling methods are illustrated in Fig. \ref{fig:app:samplingcol} for the case of the Cha II cloud column density distribution, with a very coarse 50-level sampling.

\FigAppSamplingCol{}

The choice of sampling is dependent on what one wants to achieve, and in particular on which column density range one wants to prioritise. The log and i-log spacings sample in more detail the low and high column densities respectively.  The min-max approach offers the advantage of simplicity of implementation and interpretation, but the bins can be used in a sub-optimal way, for example sampling too densely a part of the column density distribution which is not very populated, to the point of creating many empty bins. Conversely, the percentile approach yields a bin spacing that is not intuitive, but adapts directly to the structure of the data. The typical shape of column density PDFs, with a main log-normal-like peak and a long power-law-like tail (see \citet{spilker21} for a discussion of PDF shapes) also entails that the min-max approach samples in far more detail the tail than the peak of the distribution, while the opposite is true for the percentiles approach. The combination of these effects yields sampling options which put more or less emphasis on the description of the low, intermediate or high column densities. Figures \ref{fig:app:samplingpdf} and \ref{fig:app:samplingmap} illustrate the results of the sampling choice on the global volume density PDF and the peak volume density map of the Cha II cloud with a 50-level sampling, compared to the reference 2000-level i-log percentile sampling used in this study.

\FigAppSamplingPDF{}

\FigAppSamplingMaps{}

One can see that, in agreement with the effects visible in Fig. \ref{fig:app:samplingcol}, the linear and i-log min-max samplings fail to reproduce the low volume density envelope of the cloud, while the linear and log percentile samplings are unable to give detailed view of the densest regions. The log min-max and i-log percentile samplings on the other hand are able to recover reliably a broader range of volume densities.

While it is of high importance when studying star formation to have a detailed view of the high-density regions of molecular clouds, this cannot be completely at the expense of other density ranges. This is due to the increment-based nature of the volume density reconstruction: the high volume densities are not reconstructed independently, but as increments on top of a lower density structure. While a small error in the reconstruction of the lower volume densities has negligible effects on the values of the highest volume densities in a cloud (as discussed in Sect. \ref{sec:discusslimit}), a significant bias in the sampling of low-to-intermediate column densities can result in a bias in the reconstructed high volume density values -- this is for example visible for the i-log min-max sampling in Fig. \ref{fig:app:samplingpdf} and \ref{fig:app:samplingmap}, where the high volume densities, while very finely sampled, are biased by a factor of a few with respect to the reference. On the other hand, given that the volume-density reconstruction is mass-conservative, even if the sampling if the highest densities is not extremely fine, the loss of detail would affect the morphology but not the mass of the high density regions, and dense gas measurements (Sect. \ref{sec:dense}) would be largely unaffected.

In that perspective, the min-max logarithmic sampling and the i-log percentile sampling perform therefore comparably well for our purpose. In the end, our choice of the i-log percentile sampling was motivated by the wish to optimise the sampling performance without having to worry about the diversity of field sizes and column density PDF shapes, and in particular to avoid empty bins.

\section{Effect of anisotropy}
\label{app:aniso}
One of the key over-simplifications of the cloud structure model used in \coltt{} is the assumption of isotropy at all scales ($\alpha = 1$). Not only is this assumption in general not true, but the degree of isotropy or anisotropy strongly depends on the studied scale (or contour area): on projected areas above $1\pc^2$, clouds might or might not be isotropic, for areas smaller than $0.1 \times 0.1 \pc^2$, dense cores are generally considered to be quite isotropic (see \eg{} B68), and, in between, all sorts of clumps, sheets and filaments can have various degrees of isotropy, in particular high-aspect-ratio filaments of projected areas of $\sim 1\times0.1\pc^2$.

It is therefore not sufficient to fix a value of $\alpha \neq 1$ to recover the effects of anisotropic gas distribution in a molecular cloud: $\alpha$ need to vary with scale. The simplest approach is to have a fixed prescription for $\alpha$ as a function of $\mathcal{A}$. We test the effects of this approach in comparison with the isotropic computation for a well-known and highly anisotropic cloud, namely Orion\,A. It is also one of the most complex and massive clouds in our sample, which makes the volume density inference particularly challenging.

Orion\,A is famous for having a ``cometary'' shape, with a high aspect ratio in the plane of the sky that actually translated to a filamentary shape in three dimension \citep{bally08,grossschedl18,rezaei20}. The molecular gas in Orion\,A displays an elongated shape of about $40\times5$\pc{} in the plane of the sky -- and we here omit the fact that while the ``head'' of the cloud (northern part, harbouring to the Integral-Shaped Filament) lies in the plane of the sky, its ``tail'' is highly inclined, leading to an actual length of the cloud of about 70\pc{} \citep{grossschedl18}. While taking into account the variable inclination of the main structures would exemplify even further the case of a high-aspect-ratio cloud, it would also increase the complexity of the anisotropic model, so we limit ourselves to the simpler assumption of a cloud lying parallel to the plane of the sky. We thus assume that on the largest scales ($40 \times 5 = 200 \pc^2)$, the cloud is roughly cylindrical, with its depth equal to its width, despite a projected aspect ratio of $\sim 8$ (thus $\alpha = 1/\sqrt{8}$). And on small scales ($0.1\times0.1 = 0.01\pc^2$), we assume that cores are isotropic ($\alpha =1$). These two constraints are smoothly joined in log-space by the following cosine interpolation, illustrated in Fig. \ref{fig:app:alpha}:
\begin{equation}
\label{eq:app:aniso}
\alpha\left( \mathcal{A} \right) =
 \begin{cases}
    1/\sqrt{8} & \mbox{ if } \mathcal{A} \geq 200\pc^2 \\
    1 & \mbox{ if } \mathcal{A} \leq 0.01\pc^2 \\
    1 + \left(\frac{1}{\sqrt{8}} -1 \right)\cdot\frac{\cos\left(\pi\frac{\log\mathcal{A} - \log 0.01}{\log 200-\log 0.01}\right)}{2} & \mbox{otherwise}
 \end{cases}
\end{equation}

\FigAppAlpha{}

An overview of the compared outputs of the isotropic and anisotropic models for volume density can be found in Fig. \ref{fig:app:oria}; we also discuss in more details the comparison of the peak volume densities and the global volume density distributions obtained by the two different models.

\FigAppOriA{}

Since on the largest scales the reconstructed volume is smaller than in the isotropic case, the lowest reconstructed densities increments are automatically higher -- which is visible in the left panel of Fig. \ref{fig:app:rescpdf} as a shift towards higher densities of the entire lower end of the PDF. On the other hand the smallest volumes, which mostly correspond to high density regions, are essentially treated in the same way by the isotropic and anisotropic models. Reconstructing volume density from increments means that a density increase on the first increments is reflected for all higher densities, but given the shape of the volume density distribution, this has a negligible effect on the highest volume densities (as we are adding a few tens of \Htpccm{} to densities of more than $10^4 \Htpccm{}$). This is visible on the left panel of Fig. \ref{fig:app:rescpdf} where at high volume densities the PDF of the anisotropic model lies above the one of the isotropic model, but barely so. We can also see in the right panel of the same figure that the anisotropic model reconstructs less (technically, zero) of the lowest volume densities of the isotropic model; it reconstructs significantly more medium densities ($\sim 2\e{2} - 2\e{3} \Htpccm{}$) as a combined effect of accumulated excesses in low-volume density increments and of $\alpha$ still significantly lower than 1; and while it also reconstructs more high volume densities, this excess decreases in a power-law-like manner.

\FigAppRescPDF{}

\FigAppRescHist{}

\FigAppRescMaps{}

The same effect is illustrated in a different way by the comparison of peak volume densities in Fig. \ref{fig:app:rescmaps}. The maps produced by the isotropic and anisotropic models (panels 2 and 3) are remarkably similar, except for the fact that the lower end of the scale of the anisotropic map is shifted to higher densities. The difference between these two maps (panel 4) shows a trend by which at higher densities (column or peak volume), the difference is larger -- which is the effects of the accumulation of small increases in volume density increments. This difference can be as much as $10^{4} \Htpccm{}$. On the other hand we can clearly see that the ratio of these two peak volume densities (panel 5) reaches almost 1 in the highest density regions: indeed a difference of $10^{4} \Htpccm{}$ represent only a few percent, if relative to values of nearly $10^{6} \Htpccm{}$.

The same trends (difference increasing with density, ratio decreasing from $\sqrt{8}$ to 1) are illustrated in the form of joint histograms in Fig. \ref{fig:app:reschist}. In summary, while in absolute terms variations in the anisotropy parameter $\alpha$ have an impact at all scales small than the scale at which anisotropy is present, in relative terms volume density is significantly affected only by anisotropy at scales close to those corresponding to the considered range of density.
\\

The final comparison that we need to do in the scope of this paper is the question of the fraction of dense gas. Taking the same volume density threshold, we obtain an increase of 14.9\% in $f_\mathrm{dg}^\rho$ for Orion A when using this anisotropic model -- which is not as little as the peak volume densities would suggest, but is difficult to comment on given that Orion\,A is not an outlier in the density vs. SFE relation. For comparison, we also computed anisotropic model where Orion\,A is assumed to be a sheet: this time, the depth along the line of sight is assumed to be equivalent to the largest, instead of the smallest, dimension in the plane of the sky, so $\alpha = \sqrt{8}$ for the largest scales. We find this time a decrease of 13.7\% for $f_\mathrm{dg}^\rho$. Lack of knowledge about the bulk line-of-sight dimension of a cloud can thus lead to a 33\% difference in inferred dense gas fraction and thus star formation efficiency. It is however not sufficient to explain order-of-magnitude differences in SFE between apparently similar clouds, such as reported for Orion\,A and the California Cloud \citep{rezaei22}, or Orion\,A and B \citep{orkisz19}: in both cases, the distribution and isotropy of structures at intermediate scales (fragmentation and dimension of sub-clouds, fraction of mass accumulated into dense filaments) greatly matters in order to estimate properly the statistical distribution of dense gas in a cloud and its contribution to star formation. Incorporating priors and measurements on fragmentation and anisotropy for each contour at each scale is however beyond the current capabilities of the \coltt{} code.

\end{appendix}

\end{document}